%% file: main.tex
\documentclass[twoside,11pt]{article}
\usepackage{jair, rawfonts}
\usepackage{theapa}

\ShortHeadings{I Will Have Order! Optimizing Orders for Fair Reviewer Assignment}
{Payan \& Zick}
\firstpageno{1}

\input{JAIR/preamble}

\begin{document}

\title{I Will Have Order!\\Optimizing Orders for Fair Reviewer Assignment}

\author{\name Justin Payan \email jpayan@umass.edu \\
\name Yair Zick \email yzick@umass.edu \\
\addr University of Massachusetts, College of Information and Computer Sciences, \\140 Governors Drive,
Amherst, MA 01002 USA}


\maketitle

\begin{abstract}

We present fast, fair, flexible, and welfare efficient algorithms for assigning reviewers to submitted conference papers. Our approaches extend picking sequence mechanisms, standard tools from the fair allocation literature to ensure approximate envy-freeness (typically envy-freeness up to one item, or EF1). 
However, fairness often comes at the cost of decreased efficiency. To overcome this challenge, we carefully select approximately optimal picking sequence orders.    
Applying a relaxation of submodularity, $\gamma$-weak submodularity, we show our Greedy Reviewer Round Robin ($\GRRR$) approach is EF1 and yields a ${(1+\gamma)}$-approximation to the maximum welfare attainable by a round-robin picking sequence mechanism under any order. We present a weighted picking sequence mechanism called \FairSeq that targets the Weighted EF1 criterion to offer fairness in a more general setting. Using data from three conferences, we show that $\FairSeq$ runs an order of magnitude faster and provides approximate envy-freeness guarantees that are violated by existing approaches. Its simple design also makes it very flexible to new assignment constraints. $\FairSeq$ is available in the  OpenReview\footnote{\url{https://github.com/openreview/openreview-matcher}} conference management platform, giving conference organizers access to faster reviewer assignment with high welfare and envy-freeness guarantees.
\end{abstract}

\section{Introduction}
\label{sec:intro}

Peer review plays a prominent role in nearly all aspects of academia. 
It serves a number of functions -- selecting the best manuscripts, assessing originality, providing feedback, and more~\cite{mulligan2013peer}. 
Given the broad application of peer review and its significant gatekeeping role, it is imperative that this process remain as objective as possible.

One important parameter is whether papers are assigned experts who are able to appropriately review their work. 
Selecting reviewers for submitted papers is therefore a crucial first step of any reviewing process. 
In large conferences such as NeurIPS, ICML, AAAI,  and IJCAI, reviewer assignment is largely automated through systems such as the Toronto Paper Matching System (\TPMS)~\cite{charlin2013toronto}, Microsoft CMT\footnote{\url{https://cmt3.research.microsoft.com/}}, or OpenReview\footnote{\url{https://openreview.net/}}. 
Inappropriately assigned reviewers may lead to failures: misinformed decisions, reviewer disinterest, and a general mistrust of the peer-review process. 

We seek reviewer assignment algorithms that are fast, fair, flexible (to new constraints), and accurate. 
Accuracy and fairness are two very important criteria in reviewer assignment~\cite{shah2019principled}, and fair division in general \cite{brandt2016handbookfairdiv}. 
Overall assignment accuracy maintains quality standards for academic publications: we want reviewers to be an overall good match to papers.  
However, it is imperative that we do not sacrifice review quality on some papers to obtain a better overall matching. 
Even if the overall assignment is good, papers with poorly matched reviewers may be unfairly rejected or receive unhelpful feedback, causing the authors real harm. 
We thus desire algorithms which are globally accurate and fair for the papers. 
In addition, we require algorithms that run quickly, since conference organizers often compute multiple potential assignments, changing constraints or freezing some matches each time. We also need our algorithms to be simple and adaptable to new constraints; 
indeed, the organizers of major conferences often have detailed constraints such as forbidding two reviewers from the same institution to be assigned to the same paper, or requiring at least one senior reviewer per paper \cite{leyton2022matching}. 
To accomplish these goals, we consider the fair reviewer assignment problem through the lens of \emph{fair allocation}.

Our principal fairness criterion is \emph{envy}: one paper envies another paper if the other's assigned reviewers constitute a better match than its own assigned reviewers. Envy-freeness and its relaxations can be used to preclude significant disparities in reviewer-paper affinity scores. 
State-of-the-art fair reviewer assignment algorithms either maximize the minimum paper affinity score or ensure all affinity scores exceed some threshold~\cite{stelmakh2019peerreview4all,kobren2019paper}. 
Although this approach ensures all papers receive reviewers that pass some minimum quality threshold, there may still be many papers with poorly-aligned reviewers that could benefit from a small number of low-cost trades. 
Envy-based criteria, like envy-freeness up to one item (EF1), ensure that no such ``obvious faults'' exist, locally balancing the assignments.

It is generally impossible to obtain envy-free allocations for indivisible items~\cite{brandt2016handbookfairdiv}; 
thus, we focus on the relaxed criterion of envy-freeness up to one item (EF1) \cite{budish2011ef1,lipton2004approximately}. 
EF1 allocations have the property that whenever a paper $p$ has higher affinity for the reviewers of a paper $p'$, it is due to a single, high-affinity reviewer rather than a complete imbalance in outcomes. 
Maximizing welfare subject to EF1 is NP-hard and is not approximable in polynomial time~\cite{barman2019fair}. Thus, we explore methods that produce EF1 allocations with good empirical performance and partial welfare guarantees using \emph{picking sequence mechanisms}. 
In standard fair allocation settings, the well-known \emph{round-robin} (\texttt{RR}) mechanism produces EF1 allocations by setting an order of agents, and letting them select one item at a time. 
Due to constraints on reviewer selection, round-robin is not EF1 for reviewer assignment. 
We thus present a variation on classic \texttt{RR}, which we term \emph{Reviewer Round Robin} ($\RRR$). 

Round-robin mechanisms assign the same number of reviewers to each paper, but sometimes conferences require variable-sized assignments. 
For instance, two-phase reviewing processes often need to assign a variable numbers of reviewers in the second phase when some reviewers failed to respond in the first phase. 
We therefore also study a family of weighted picking sequences which satisfy the Weighted EF1 constraint (WEF1)~\cite{wef1}. 
This constraint generalizes EF1 to settings where agents have unequal item demands, ensuring that agents are EF1 after normalizing by their demands. 
Our algorithms, Weighted Reviewer Picking ($\WRP$) and $\FairSeq$, follow $\RRR$ by applying a standard weighted picking sequence with some added steps to accommodate the constraints of reviewer assignment. 

While picking sequence mechanisms are known to satisfy fairness constraints, their efficiency guarantees are highly dependent on the order in which players pick items. For example, consider a stylized setting where there are two papers ($p_i$ and $p_j$) and two reviewers ($r_1$ and $r_2$): paper $p_i$ has an affinity score of 5 for both reviewers, while paper $p_j$ has a score of 10 for $r_1$ and 0 for $r_2$. A round-robin mechanism that lets $p_i$ pick first runs the risk of having $p_i$ pick $r_1$, leaving $p_j$ with $r_2$. Letting $p_j$ pick first results in a much better outcome, without compromising on fairness. It is generally difficult to identify optimal picking sequences~\cite{bouveret2011general,kalinowski2013social,aziz2015possible,aziz2016welfare}. 
We ask the question: Can we identify \emph{approximately optimal} paper orders?

\subsection{Our Contributions}\label{sec:contrib}
We run a combinatorial search for orders of papers that yield high efficiency allocations for picking-sequence mechanisms like $\RRR$ and $\WRP$. 
To this end, we examine the problem of finding an optimal paper order via the lens of \emph{submodular optimization}. 
We optimize a function on partial paper sequences, which varies according to the welfare of the allocation resulting from the picking sequence. 
This function is not submodular in general, but we can capture its distance from submodularity via a variable $\gamma$. 
Our main theoretical result (Theorem \ref{thm:main_result}), which is of independent interest to the fair division community, shows that a simple greedy approach maximizes this function up to a factor of $1+\gamma$. 
We call this approach Greedy Reviewer Round Robin ($\GRRR$). 
The approach can also be applied to weighted picking sequences like $\WRP$, by optimizing over the order in which ties in priority are broken. 

Though we do not offer theoretical welfare guarantees for the $\FairSeq$ algorithm, we present a heuristic approach that optimizes for a high welfare weighted picking sequence.  $\FairSeq$ breaks ties in priority order adaptively when they occur, by assigning a reviewer to any paper with top priority which can receive the highest increase in welfare. This approach is  fast and straightforward to implement. 
\FairSeq thus achieves our four goals: it is a fast, fair, flexible, and welfare efficient reviewer assignment mechanism. 


We compare our $\GRRR$ and $\FairSeq$ algorithms with three other state-of-the-art paper assignment frameworks on three real-world conference datasets.

Not only are $\GRRR$ and $\FairSeq$ the only provably (W)EF1 approaches, $\FairSeq$ is an order of magnitude faster than the other methods on all datasets. Most importantly, we show that $\FairSeq$ is often significantly more fair than \FairFlow, a fair allocation protocol \cite{kobren2019paper}, on nearly all fairness metrics. 

This work extends a previous conference proceedings paper \cite{payan2022will}. 
We have added algorithms that satisfy variable paper demands and drastically improve on runtime ($\WRP$ and $\FairSeq$), as well as more detailed experimental analysis.

\subsection{Related work}
\label{sec:related_work}

The reviewer assignment problem has been widely studied. Most works model the problem as a mixed-integer linear program maximizing affinity between reviewers and papers; the Toronto Paper Matching System (\TPMS) is the most notable work with this formulation~\shortcite{charlin2011framework,charlin2013toronto}. The affinity typically models alignment between reviewer expertise and paper topics, but can incorporate other relevant notions like reviewer bids, conflicts of interest, and author suggestions; many works study how these values are generated and are orthogonal to our work \shortcite{leyton2022matching,mysore2023editable,rozenzweig2023mitigating}. Affinities are generally considered a good proxy for value at both an individual and collective level, since higher-affinity reviewers will typically be more qualified for and interested in a paper, resulting in more detailed and accurate reviews.
Affinity scores are universally available in systems like \TPMS, Microsoft CMT, or OpenReview, and it is standard practice to use these affinity scores to compute welfare and fairness measures~\shortcite{stelmakh2019peerreview4all,kobren2019paper,jecmen2020mitigating,leyton2022matching,shah2022challenges,cousins2023into}.

A number of prior works consider fairness objectives in peer review, though none of them consider envy-freeness up to one item. \shortciteA{hartvigsen1999conference} ensure that at least one qualified reviewer is assigned to each paper. A few recent approaches approximately maximize the minimum paper score, or maximize the sum of scores subject to a minimum individual score threshold~\cite{odell2005paper,kobren2019paper,stelmakh2019peerreview4all}. \shortciteA{aziz2023group} present a reviewer assignment algorithm that satisfies the \emph{core}, ensuring that no group of papers can deviate without reducing their total welfare.
A number of works study fair assignment of papers to reviewers, allowing reviewers to express preferences over papers by bidding~\cite{garg2010assigning,lian2018conference,aziz2019constrained,tan2021minimal}. This setting aims to be fair to the reviewers rather than the papers, as is the case for our work.\footnote{We briefly discuss fairness to reviewers in \Cref{appx:rev_loads}.} 
Other works target reviewer assignments with properties besides fairness or efficiency; \shortciteA{long2013good} avoid conflicts of interest, while \shortciteA{kou2015weighted} and \shortciteA{ahmed2017diverse} assign sets of reviewers with diverse interests and full coverage of the papers' topics. 


A few works in fair allocation are relevant for fair reviewer assignment, but they have important limitations in our setting. \shortciteA{aziz2019constrained} present an algorithm which attempts to output a $W$-satisfying EF1 allocation for constraint(s) $W$. When $W$ includes a minimum threshold for welfare, their approach is somewhat similar to ours. However, rather than greedily maximizing welfare by letting the locally optimal agent pick, they let any agent pick as long as $W$ can still be achieved. 
\shortciteA{biswas2018fair} present a modification of the round-robin mechanism that assigns a complete EF1 allocation when items are partitioned into categories and agents can receive a limited number of items from each category, but the overall number of items per agent is unlimited. \shortciteA{dror2021fair} study fair allocation under matroid constraints, but only for identical or binary valuations, less than four agents, or a single uniform matroid constraint. \shortciteA{caragiannis2023optimizing} show that greedily selecting agents is a $2$-approximation for the maximum welfare picking sequence when agents can choose at most one item.

Weighted envy-freeness up to one item was one of the first fairness notions studied for agents with unequal entitlements \cite{benabbou2020finding}. Much recent work in fair allocation has focused on this and other guarantees to agents with unequal entitlements~\cite{babaioff2021fair,chakraborty2022weighted,suksompong2022maximum,suksompong2022maximum,SUKSOMPONG202348,montanari2024weighted}. Much of this work focuses on the setting of binary valuations or binary submodular valuations. Of particular relevance is recent work recommending usage of picking sequences for weighted fair division \cite{CHAKRABORTY2021103578}. 
 
Our application of submodular optimization to optimizing orders for round-robin is inspired by previous work on fair allocation with submodular \emph{valuations}~\cite{benabbou2020finding,babaioff2020fair,barman2021existence,viswanathan2023yankee}.
Prior work has also studied maximization of approximately submodular functions, though none has combined matroid constraints with a definition of approximate submodularity similar to ours~\cite{das2011submodular,golz2019migration}.

Existing work shows the hardness of maximizing welfare for EF1 and picking sequence  allocations. \shortciteA{aziz2019constrained} show maximizing welfare subject to EF1 is NP-hard, and \shortciteA{barman2019fair} show the same problem is not even polynomial-time approximable. \shortciteA{aziz2016welfare} show that the problem of determining if a given welfare is possible under a picking sequence of a certain class is NP-complete for some classes of picking sequences (but not round-robin).

Finally, our $\FairSeq$ algorithm uses a subroutine inspired by the \emph{exchange graph} routine from the Yankee Swap algorithm \cite{viswanathan2023general,viswanathan2023yankee,cousins2023dividing,cousins2023good}.


\section{Preliminaries}
\label{sec:problem}
We represent reviewer assignment as a problem of allocating indivisible goods, 
with papers as agents and reviewers as goods. 
We are given a set of $n$ papers $P = \{p_1,\dots,p_n\}$, and a set of $m$ reviewers $R = \{r_1, r_2, \dots ,r_m\}$. 
Each paper $p$ has an affinity function over reviewers $v_p :R \to \mathbb{R}_{\geq 0}$, which defines the suitability of the reviewer for the paper.

Papers generally receive more than one reviewer, so we define affinity functions over sets of reviewers. We assume \emph{additive} functions, where for a paper $p \in P$ and a subset $S \subseteq R$, $v_p(S) = \sum_{r \in S} v_p(r)$. An \emph{assignment} or \emph{allocation} of reviewers to papers is an ordered tuple $A = (A_{p_1}, A_{p_2}, \dots A_{p_n})$ where each $A_p \subseteq R$ is a set of distinct reviewers assigned to paper $p$. 
Each paper $p$ requires $k_p$ reviewers. Often we have a fixed $k$ such that $k_p = k$ for all $p$. An allocation is \emph{complete} if every paper $p$ is assigned $k_p$ distinct reviewers (and \emph{incomplete} otherwise). We refer to $A_p$ as paper $p$'s \emph{bundle}, and $v_p(A_p)$ as paper $p$'s \emph{valuation} under $A$. Given a paper $p_i$, we will sometimes simplify notation and write $v_i$ and $A_i$ instead of $v_{p_i}$ and $A_{p_i}$.

Each reviewer $r \in R$ has an upper bound $u_r$ on the number of papers they can review, and may have lower bounds $l_r$, which help ensure a more even distribution of work. Given an allocation $A$, we denote the number of papers to which a reviewer $r$ is assigned as $c_r^{A} = \sum_{p \in P} |\{r\} \cap A_p|$, or just $c_r$ when $A$ is clear from context.  

Often, there are additional constraints on the reviewer assignment. We model these additional constraints as a function $C$, where $C(p, r, A) \in \{True, False\}$ indicates whether reviewer $r$ can be assigned to paper $p$ given the incomplete allocation $A$. We will require these constraints to be monotone; if we have two allocations $A=(A_1, A_2, \dots, A_n)$ and $A'=(A'_1, A'_2, \dots, A'_n)$ with $A_p \subseteq A'_p$ for all $p$, then $C(p, r, A)$ implies $C(p, r, A')$. Conflicts of interest represent the simplest such constraints; $C(p, r, A) = True$ when there is a conflict of interest between paper $p$ and reviewer $r$. Since $C(p, r, A)$ does not depend on $A$ for conflicts of interest, we will simply write $C(p, r)$ when no additional constraints apply. We sometimes apply more detailed constraints both within paper bundles (e.g., no paper can receive two reviewers from the same institution) or across multiple bundles (e.g., an anti-collusion constraint requiring that there is no pair of papers $p$ and $p'$ and reviewers $r$ and $r'$ such that $r$ authored $p$ and requested to review $p'$, $r'$ authored $p'$ and requested to review $p$, $r$ is assigned to $p'$, and $r'$ is assigned to $p$). These complex constraints are increasingly used by large conferences \cite{leyton2022matching}.

To simplify notation, given a set $X$ and an element $y$, we often write $X+y$ and $X-y$ instead of $X \cup\{y\}$ and $X \setminus \{y\}$. 

We now discuss our notion of fairness. An allocation $A$ is considered \emph{envy-free} if for all pairs of papers $p$ and $p'$, $  v_p(A_p) \geq v_p(A_{p'})$. This criterion is not achievable in general (consider the example of two papers and one reviewer $r$ whose upper bound is $u_r=1$), so we relax the criterion. An allocation $A$ is envy-free up to one item (EF1) if for all pairs of papers $p$ and $p'$, either $v_p(A_p) \geq v_p(A_{p'})$ or $\exists r \in A_{p'}$ such that $v_p(A_p) \geq v_p(A_{p'} - r) $. EF1 should only be used when all papers have the same demands and $k_p = k$ for all $p$. When papers have distinct demands $k_p$, we apply the fairness notion of \emph{weighted} envy-freeness~\cite{wef1}. 
An allocation $A$ is considered weighted envy-free if for all pairs of papers $p$ and ${p'}$, $ \frac{v_p(A_p)}{k_p} \geq \frac{v_p(A_{p'})}{k_{p'}} $. Analogously, an allocation $A$ is weighted envy-free up to one item (WEF1) if for all pairs of papers $p$ and ${p'}$, either $\frac{v_p(A_p)}{k_p} \geq \frac{v_p(A_{p'})}{k_{p'}} $ or there is a reviewer $r \in A_{p'}$ such that $\frac{v_p(A_p)}{k_p} \geq \frac{v_p(A_{p'} - r)}{k_{p'}} $.

The \emph{utilitarian social welfare} (also called utilitarian welfare or USW) of an allocation $A$ is the sum of the papers' valuations under that allocation.
$\USW$ is a natural objective in the context of reviewer assignment, and has been used in many prior works on this topic~\cite{conry2009recommender,charlin2013toronto,kobren2019paper,stelmakh2019peerreview4all,payan2022will,cousins2023into}. 


For picking sequences, we define an \emph{order} on papers $p \in P$ as a tuple $\mathcal{O} = (S, o)$, where $S \subseteq P$ is the set of papers in the order and $o : S \to \{1, 2, \dots , |S|\}$ is a permutation on $S$ mapping papers to positions. Let $\Omega(P)$ denote the set of all orders over subsets of $P$. We will also refer to $\Omega(R)$, the set of all orders over subsets of $R$, defined analogously to $\Omega(P)$. We slightly abuse notation and say that a paper $p \in \mathcal{O}$ if $p \in S$. 
For any $p, p' \in \mathcal{O}$, we say that $p \prec_\mathcal{O} p'$ if and only if $o(p) < o(p')$. 
We write $p \prec p'$ when $\cal O$ is clear from context. We can think of an order $\cal O = (S, o)$ as an ordered list $[o_1, o_2, \dots o_{|S|}]$ such that $o_l=o^{-1}(l)$ for all positions $l$. We use the notation $\mathcal{O} + p$ to indicate the order $(S', o')$ that appends $p$ to the end of $\mathcal{O}$. Formally, $S' = S + p$, $o'(p') = o(p')$ for $p' \in S$, and $o'(p) = |S'|$. We write the empty order as $\cal O_\emptyset = (\emptyset, o)$ with any function $o$.

\section{Fair and Efficient  Assignment with Reviewer Round Robin}
\label{sec:rr_allocs}



We first show how to obtain EF1 reviewer assignments when all papers have equal demands $k_p = k$, and reviewers do not have lower bounds $l_r$; we handle the more general case in \Cref{sec:wef1}.
Our algorithm draws upon the simple and well-known \emph{round-robin} mechanism. 
Given an ordered list of papers, round-robin proceeds in rounds. In each round, we iterate over the papers in the provided order, assigning each paper its highest valued remaining reviewer. This allocation is EF1 for additive valuations without uniqueness constraints by a simple argument~\cite{caragiannis2019unreasonable}. 
For any paper $p$, we divide the assignments into rounds. 
Paper $p$ prefers its own reviewers to the reviewers of any paper $p' \succ p$, and it prefers its own bundle to that of any agent $p' \prec p$ if we ignore the reviewer given to $p'$ in the first round.

The constraint that papers must be assigned at most $k$ \emph{distinct} reviewers poses a challenge. 
A trivial modification of round-robin allows us to satisfy the cardinality constraint --- proceed for exactly $k$ rounds, then stop. We might naively update round-robin to satisfy the distinctness constraint as well, by assigning each paper the best reviewer they do not already have. 
However, the argument from~\citeA{caragiannis2019unreasonable} fails. To see why, suppose a paper $p$ is assigned a reviewer $r$ in one round. In the next round, $p$ may still prefer $r$ over any other reviewer, but we cannot assign it. We will be forced to assign $p$ a much worse reviewer, giving another paper that desired ``second copy'' of $r$. A more detailed counterexample is shown in Table~\ref{tab:rr_not_ef1}.

\begin{table}[ht!]
        \caption{The naive constrained round-robin, where each paper is assigned the best reviewer they do not already have, fails for the example valuations shown below on 4 papers and 6 reviewers. Suppose $u_r=2$ for all $r$ and $k = 3$. If we apply naive constrained round-robin with the papers in increasing numerical order, we obtain the allocation $(A_{1} = \{r_1, r_5, r_6\}, A_{2}= \{r_4, r_1, r_3\}, A_{3}= \{r_4, r_5, r_2\}, A_{4}= \{r_3, r_2, r_6\})$. However, $v_4(A_{2} - r) \geq 5$ for all $r \in A_{2}$, while $v_4(A_{4}) = 4+\epsilon$. In contrast, the allocation $(A_1 = \{r_1, r_5, r_6\}, A_2= \{r_4, r_1, r_2\}, A_3= \{r_4, r_5, r_3\}, A_4= \{r_3, r_2, r_6\})$ is EF1.}
        \begin{tabu} to \columnwidth {XXXXXXX}
            & $r_1$ & $r_2$ & $r_3$ & $r_4$ & $r_5$ & $r_6$ \\
            \midrule
            $p_1$ & $2$ & $0$ & $0$ & $1$ & $0.5$ & $\epsilon$ \\
            $p_2$ & $3$ & $1$ & $2$ & $10$ & $0$ & $0$ \\
            $p_3$ & $0$ & $\epsilon$ & $0$ & $10$ & $1$ & $0$ \\
            $p_4$ & $2$ & $1$ & $3$ & $10$ & $0$ & $\epsilon$ \\

        \end{tabu}
        \label{tab:rr_not_ef1}
\end{table}


We present a modification of round-robin that takes a paper order $\cal O$ and assigns reviewers to papers in order $\cal O$ such that all constraints are satisfied and the allocation is EF1. Reviewer Round Robin or \RRR~(Algorithm~\ref{alg:rrr}) forbids any assignment that violates a crucial invariant for proving EF1. This invariant derives from the proof of EF1 in the additive case. Any time we would assign a reviewer such that EF1 would be violated, we forbid the assignment and instead assign a different reviewer. 
EF1 violations can only arise when another paper preferred that reviewer but the assignment was forbidden, either because it had been assigned already, or because it would have caused an EF1 violation for that paper as well. 
We always attempt to assign reviewers in preference order. Thus when we attempt to assign a reviewer $r$ to paper $p$, we only need to check for EF1 violations against other papers to which we have attempted to assign $r$ in the past. 
Theorem~\ref{thm:RRR_ef1} asserts the correctness of \RRR.

\begin{algorithm}[t]
    \caption{Reviewer Round Robin ($\RRR$)}
    \begin{algorithmic}[1]
        \REQUIRE {Reviewers $R$, reviewer upper limits $u_r$, paper order $\cal O$, affinity functions $v_p$, bundle size limit $k$, constraints $C$}

        \STATE Initialize allocation $A$ as $A_p \gets \emptyset$ for all papers $p \in \cal{O}$
        \STATE Initialize the attempted set $S_p \gets \emptyset$ for all  $p$
        
        \FOR{Round $t \in \{1, \dots, k\}$}
            \FOR{$p \in \cal O$ in order}
               
                \FOR{Reviewer $r$ in decreasing order of $v_p(r)$ (break ties lexicographically)}
                    \STATE Attempt to assign $r$ to $p$ ($S_p \gets S_p + r$)
                    \IF {$c_r^{A} < u_r$, $r \notin A_p$, and $\neg C(p,r,A)$}
                        
                        \IF{No $p'$ with $r \in S_{p'}$ envies $A_p + r$ more than $1$ reviewer}
                            \STATE $A_p \gets (A_p + r)$
                            \STATE Move to the next paper in $\cal O$
                        \ENDIF
                    \ENDIF
                \ENDFOR
                \STATE If no new reviewer is assigned to $p$, return $A$

            \ENDFOR
        \ENDFOR
        \RETURN $A$
    \end{algorithmic}
    \label{alg:rrr}
  \end{algorithm}
  
\begin{theorem}
    \label{thm:RRR_ef1}
    \RRR~terminates with an EF1 allocation where papers receive at most $k$ distinct reviewers, no reviewer $r$ is assigned to more than $u_r$ papers, and all constraints $C$ are satisfied.
\end{theorem}

\begin{proof}
    The algorithm assigns at most one reviewer to each paper in each round for $k$ rounds, so the constraint that all papers receive at most $k$ reviewers is satisfied. 
    In addition, the algorithm always checks that $r \notin A_p$, the number of papers which already have $r$ is no more than $u_r$, and $C(p, r, A)$ is satisfied before assigning $r$ to $p$. 
    Thus no paper receives duplicate reviewers, reviewer upper bounds are satisfied, and additional constraints $C$ are never violated.

    We now prove that the returned allocation is EF1. 
    Consider some arbitrary paper ${p'}$; 
    we show that ${p'}$ does not envy any other paper by more than 1 reviewer. As in the original round-robin argument, we divide the assignments of reviewers to papers into rounds
    $0, 1, \dots, s$, where $s \le k$ ($s < k$ only when the algorithm terminates early). 
    Round 
    $0$ contains the assignments made during iteration 1 of Algorithm~\ref{alg:rrr} to all papers preceding $p'$ in $\cal O$. Rounds 
    $1$ through $s-1$ begin with the assignment of a new reviewer to ${p'}$ and end with the assignment of a new reviewer to the paper immediately preceding $p'$ in $\cal O$, while round 
    $s$ begins with assignment to ${p'}$ and ends with assignment to some paper after ${p'}$.

    Consider the bundle $A_p^{(t)}$ assigned to some paper $p \neq p'$ after the end of some round 
    $t\in \{0, 1, \dots s\}$. 
    We will define modified bundles $B_p^{t}$ for all $p$, and prove by induction that $v_{p'}\left(B_p^{(t)}\right) \leq v_{p'}\left(A_{p'}^{(t)}\right)$. For all $t$, let $B_p^{(t)} = A_p^{(t)}$ if $p' \prec p$. If $p' \succ p$, $B_p^{(t)} = A_p^{(t)} - \argmax_{r \in A_p^{(t)}}v_{p'}(r)$ .

    For the base case, we see that after round 
    $0$, 
    $\abs{B_p^{(0)}} = 0$
    for all $p$ and $\abs{A_{p'}^{(0)}} = 0$, so $v_{p'}\left(A_{p'}^{(0)}\right) = v_{p'}\left(B_p^{(0)}\right) = 0$.
    
    Now suppose that after round $t-1$, we have $v_{p'}\left(A_{p'}^{(t-1)}\right) \geq v_{p'}\left(B_p^{(t-1)}\right)$ for all $p$. Suppose there is some $p$ such that after round $t$, $v_{p'}\left(B_p^{(t)}\right) > v_{p'}\left(A_{p'}^{(t)}\right)$. ${p'}$ is assigned a reviewer in all rounds except $0$, and because affinities are non-negative $p$ must be assigned a reviewer in round $t$ to obtain $v_{p'}\left(B_p^{(t)}\right) > v_{p'}\left(A_{p'}^{(t)}\right)$. By the inductive hypothesis and the fact that affinities are additive, ${p'}$ must prefer the reviewer $p$ was assigned in $t$, $r_p$, to the reviewer ${p'}$ was assigned in $t$, $r_{p'}$. Because ${p'}$ went first in $t$, this means that we attempted to assign $r_p$ to ${p'}$ either in $t$ or earlier and $p$ must have checked for envy against ${p'}$. This is a contradiction, since $v_{p'}\left(B_p^{(t)}\right) > v_{p'}\left(A_{p'}^{(t)}\right)$ violates EF1. 

    
\end{proof}

It is possible for \RRR~to return an incomplete allocation when a complete one exists. This is tight in some sense; we can easily show an instance of the reviewer assignment problem where no valid, complete allocation is EF1. Consider the case with two papers $P = \{p_1, p_2\}$ with $k_p=2$ for all $p \in P$ and four reviewers $R = \{r_1, r_2, r_3, r_4\}$ with $u_r = 1$ for all $r \in R$. Suppose that $v_1(r_1) = v_1(r_2) = 1$, $v_1(r_3) = v_1(r_4) = 0$, and $C(p_1, r_1, A) = C(p_1, r_2, A) = \textit{True}$ for all $A$. Then the only valid complete allocation is $A = \{\{r_3, r_4\}, \{r_1, r_2\}\}$, which is not EF1.

It is straightforward to show that $\RRR$ always returns a complete, EF1 allocation when the number of reviewers is large and there are no additional constraints $C$ (the proof of \Cref{prop:large_revs} is given in \Cref{appx:missing proofs}).

\begin{restatable}{proposition}{largerevs}
    \label{prop:large_revs}
    Given a reviewer assignment problem with $m$ reviewers, $n$ papers, no constraints $C$, and $k$ paper bundle size limits, where $m \geq kn$, \RRR~returns a complete and EF1 allocation.
\end{restatable}



\section{Non-Uniform Demands and Minimum Reviewer Supply}
\label{sec:wef1}

Papers may sometimes require different numbers of reviewers; conference organizers may run reviewer assignment multiple times to account for late reviews, borderline papers, and other mitigating circumstances. In addition, conference organizers might wish to require that each reviewer receives a minimum number of papers to review. These reviewer lower bounds ensure more balanced workloads for the reviewers. To satisfy these additional real-world constraints, we introduce variants of $\RRR$ that allow for variable paper demands $k_p$ and reviewer lower bounds $l_r$. 

To accommodate variable paper demands, we present a weighted analogue of $\RRR$ called Weighted Reviewer Picking ($\WRP$). We replace round-robin with the picking sequence introduced by~\citeA{wef1}, which guarantees WEF1. Papers no longer receive assignments in a fixed order; at each iteration the paper which has reached the smallest fraction of its bundle size limit $k_p$ is chosen to receive the next reviewer. We break ties in this ``fraction of satisfied demand'' criterion by consulting a fixed tie-breaking order $\cal O$. Algorithm~\ref{alg:wrrr} shows the complete approach.

\begin{algorithm}[t]
    \caption{Weighted Reviewer Picking ($\WRP$)}
    \begin{algorithmic}[1]
        \REQUIRE {Reviewers $R$, reviewer upper limits $u_r$, paper order $\cal O$, affinity functions $v_p$, bundle size limits $k_p$, constraints $C$}

        \STATE Initialize allocation $A$ as $A_p \gets \emptyset$ for all papers $p \in \cal O$
        \STATE Initialize the attempted set $S_p \gets \emptyset$ for all  $p$
        
        \WHILE{$\exists p: |A_p| < k_p$}
            \STATE ${p^\ast} \gets \argmin_{p \in P}\frac{|A_p|}{k_p}$, breaking ties using $\cal O$

            \FOR{Reviewer $r$ in decreasing order of $v_{p^\ast}(r)$ (break ties lexicographically)}
            \STATE Attempt to assign $r$ to ${p^\ast}$ ($S_{p^\ast} \gets S_{p^\ast} + r$)
                \IF {$c_r^{A} < u_r$, $r \notin A_{p^\ast}$, and $\neg C({p^\ast}, r, A)$}
                    
                    \IF{All $p' \neq {p^\ast}$ with $|S_{p^\ast} \cap S_{p'}| > 0$ satisfy WEF1 with respect to ${p^\ast}$}
                        \STATE $A_{p^\ast} \gets (A_{p^\ast} + r)$
                        \STATE Move to the next paper in $\argmin_{p \in P}\frac{|A_p|}{k_p}$
                    \ENDIF
                \ENDIF
            \ENDFOR
            \STATE If no new reviewer is assigned to ${p^\ast}$, return $A$
            
        \ENDWHILE
        \RETURN $A$
    \end{algorithmic}
    \label{alg:wrrr}
  \end{algorithm}
Parallel to Theorem~\ref{thm:RRR_ef1}, we state that Algorithm~\ref{alg:wrrr} is WEF1.

\begin{theorem}
    \label{thm:WRRR_wef1}
    Algorithm~\ref{alg:wrrr} terminates with a WEF1 allocation where papers receive at most $k_p$ distinct reviewers, no reviewer $r$ is assigned to more than $u_r$ papers, and additional constraints $C$ are satisfied.
\end{theorem}

\begin{proof}
    As long as there is some paper $p$ with $|A_p| < k_p$, we will never pick a paper $p'$ with $|A_{p'}| = k_{p'}$ in the picking sequence, since $\frac{|A_p|}{k_p} < 1 = \frac{|A_{p'}|}{k_{p'}}$. This proves that when we terminate, no papers have more than $k_p$ reviewers.
 
    In addition, the algorithm always checks that $r \notin A_p$, $c_r^A < u_r$, and $\neg C(p, r, A)$ before assigning $r$ to $p$. 
    Thus no paper receives duplicate reviewers, reviewer upper bounds are satisfied, and additional constraints $C$ are satisfied.
 
    We now show that the allocation is WEF1, by showing that after a paper $p$ is assigned a reviewer $r$, all papers are WEF1 with respect to $p$. Assuming all $k_p \geq 1$, the first $n$ iterations of the algorithm assign each paper a single reviewer, in order of $\mathcal{O}$. The allocation is clearly WEF1 at each of those iterations. Suppose that now $p$ has been assigned a reviewer $r$, after all papers have at least one reviewer. For any $p'$ which we have attempted to give a reviewer that we have also attempted to give to $p$ ($|S_p \cap S_{p'}| > 0$), we have already checked that ${p'}$ does not have weighted envy for $p$ over one item. 
    
    It remains to analyze the case when $|S_{p} \cap S_{p'}| = 0$. We have not attempted to give ${p'}$ any reviewers that we have also attempted to give to $p$, including $r$. Thus, for any reviewer $r_{p'}$ given to ${p'}$, $v_{p'}(r_{p'}) > v_{p'}(r_p)$ for any $r_p$ that $p$ was given after ${p'}$ was given $r_{p'}$. This criterion is sufficient for the proof from~\cite{wef1} to go through.
 \end{proof}

\subsection{Generalized Reviewer Picking Algorithms}

Both $\RRR$ and $\WRP$ can be viewed as instantiations of a broader meta-algorithm, depicted in \Cref{alg:pickingreviewers}. This algorithm applies a paper selection criterion $\PapCrit$, and a reviewer selection criterion $\RevCrit$. The paper selection criterion $\PapCrit(A, \{k_p\}_{p \in P}, \{l_r\}_{r \in R}, \{u_r\}_{r \in R}) \in \Omega(P)$ computes, given a partial allocation $A$, an ordered set of papers that can select a reviewer. $\PapCrit$ may also take as an argument a fixed order over papers, $\cal O$. If it does, then the paper selection function will simply select the next paper in the order $\cal O$. Then, the reviewer selection criterion $\RevCrit(p, A, \{k_p\}_{p \in P}, \{l_r\}_{r \in R}, \{u_r\}_{r \in R}, C) \subseteq \Omega(R)$ computes a ordered set of reviewers that paper $p$ can be assigned. Since most of the arguments of these criteria are clear from context, we will write simply $\PapCrit(A(, \cal O))$ and $\RevCrit(A, p)$. 

Concretely, $\RRR$ can be implemented by requiring $\PapCrit(A, \cal O)$ to select the singleton set containing the next paper in the order $\cal O$, and requiring $\RevCrit(A, p)$ to select reviewers $r$ such that $c_r^A < u_r$, $r \notin A_p$, $\neg C(p, r, A)$, and no other paper $p'$ would envy $p$ more than $1$ item after adding $r$ to $A_p$. Naturally, $\RevCrit(A, p)$ is ordered in decreasing order of $v_p(r)$. $\WRP$ is implemented with $\PapCrit(A) = \argmin_{p \in P} \frac{|A_p|}{k_p}$, ordered by $\cal O$, and $\RevCrit(A, p)$ defined analogously to before. $\RevCrit(A, p)$ now requires WEF1 to be satisfied, rather than EF1.

\begin{algorithm}[t]
    \caption{Picking Sequence Reviewer Assignment}
    \begin{algorithmic}[1]
        \REQUIRE {$m$ reviewers $R$, reviewer upper limits $u_r$, $n$ papers $P$, affinity functions $v_p$, bundle size limits $k_p$, constraints $C$, paper selection criterion $\PapCrit$, reviewer selection criterion $\RevCrit$}
        \STATE Initialize allocation $A$ as $A_p \gets \emptyset$ for all papers $p \in P$
        \WHILE{$A$ is not complete}
            \IF{$\PapCrit(A) \neq \cal O_\emptyset$ and $\RevCrit(A, p) \neq \cal O_\emptyset$ for some $p \in \PapCrit(A)$}
                \STATE $p \gets $ first paper in $\PapCrit(A)$ with $\RevCrit(A, p) \neq \cal O_\emptyset$
                \STATE $r \gets $ first reviewer in $\RevCrit(A, p)$
                \STATE $A_p \gets A_p + r$
            \ELSE
                \RETURN $A$
            \ENDIF
        \ENDWHILE

        \RETURN $A$
    \end{algorithmic}
    \label{alg:pickingreviewers}
  \end{algorithm}
\subsection{Minimum Reviewer Supply}

We can also easily introduce minimum reviewer supply constraints. We will utilize a simple trick to satisfy these constraints, which can be applied to any instantiation of \Cref{alg:pickingreviewers}. When the remaining demand equals the number of assignments required to meet reviewer minima, we restrict the available set of reviewers to those who need to be assigned to meet minimum requirements. More formally, when we have 

\begin{align}
    \sum\limits_{r \in R} \max(l_r - c_r^{A}, 0) = \sum\limits_{p \in P} k_p - |A_p| , \notag
\end{align}
we set $u_r = \max(l_r - c_r^A, 0)$ for all reviewers $r$. Thus, we spend the final $\sum_{p \in P} k_p - |A_p|$ steps assigning exactly the reviewers needed to meet reviewer minima. If the algorithm terminates with a complete allocation, it will satisfy the reviewer lower bounds. 

A more general version of this modification takes as input a bound modification function $h$, such that $h(A, \{k_p\}_{p \in P}, \{l_r\}_{r \in R}, \{u_r\}_{r \in R})$ takes the current allocation $A$ and problem-specific reviewer and paper bounds and determines if the bounds need to be modified to satisfy a constraint. The modified meta-algorithm is displayed in \Cref{alg:pickingreviewerslowerbounds}.



Theorems~\ref{thm:rrr_revmin} and~\ref{thm:wrrr_revmin} parallel Theorems~\ref{thm:RRR_ef1} and~\ref{thm:WRRR_wef1}, showing that when we modify $\RRR$ and $\WRP$ to include the bound modification function $h$, they return EF1 (WEF1) allocations satisfying all constraints.

\begin{algorithm}[t]
\caption{Picking Sequence Reviewer Assignment with Reviewer Lower Bounds}
\begin{algorithmic}[1]
    \REQUIRE {$m$ reviewers $R$, reviewer upper limits $u_r$,  reviewer lower limits $l_r$, $n$ papers $P$, affinity functions $v_p$, bundle size limits $k_p$, constraints $C$, paper selection criterion $\PapCrit$, reviewer selection criterion $\RevCrit$, bound modification function $h$}
    \STATE Initialize allocation $A$ as $A_p \gets \emptyset$ for all papers $p \in P$
    \WHILE{$A$ is not complete}
        \STATE $\{k_p\}_{p \in P}, \{l_r\}_{r \in R}, \{u_r\}_{r \in R} \gets h(A, \{k_p\}_{p \in P}, \{l_r\}_{r \in R}, \{u_r\}_{r \in R})$
        \IF{$\PapCrit(A) \neq \cal O_\emptyset$ and $\RevCrit(A, p) \neq \cal O_\emptyset$ for some $p \in \PapCrit(A)$}
            \STATE $p \gets $ first paper in $\PapCrit(A)$ with $\RevCrit(A, p) \neq \cal O_\emptyset$
            \STATE $r \gets $ first reviewer in $\RevCrit(A, p)$
            \STATE $A_p \gets A_p + r$
        \ELSE
            \RETURN $A$
        \ENDIF
    \ENDWHILE

    \RETURN $A$
\end{algorithmic}
\label{alg:pickingreviewerslowerbounds}
\end{algorithm}

\begin{theorem}
    \label{thm:rrr_revmin}
    \Cref{alg:pickingreviewerslowerbounds}, implemented using the $\PapCrit$ and $\RevCrit$ from $\RRR$, terminates with an EF1 allocation where papers receive at most $k$ distinct reviewers, all constraints $C$ are satisfied, and no reviewer $r$ is assigned to more than $u_r$ papers. If the algorithm assigns $k$ reviewers to each paper, then all reviewers will be assigned to at least $l_r$ papers.
\end{theorem}

\begin{proof}
   The distinctness of reviewers per paper, reviewer upper bounds $u_r$, satisfaction of $C$, and upper limit on $k$ reviewers per paper are satisfied for the same reasons described in Theorem~\ref{thm:RRR_ef1}. 

   We show that when the algorithm terminates with a complete allocation, the reviewer minima have been satisfied. We assume that the assignment problem is feasible ($\sum_{p \in P} k_p \geq \sum_{r \in R} l_r$), so  $\sum_{p \in P} k_p - |A_p| \geq \sum_{r \in R} \max(l_r - c_r^{A}, 0)$ at the beginning of the assignment process. $\sum_{p \in P} k_p - |A_p|$ decreases by $1$ each time a paper is assigned a reviewer and reaches $0$ by the end of the assignment process, and $\sum_{r \in R} \max(l_r - c_r^{A}, 0)$ decreases by either $1$ or $0$ each iteration. At some point, either $\sum_{r \in R} \max(l_r - c_r^{A}, 0) = 0$ and thus all lower bounds are satisfied, or we have $\sum_{r \in R} \max(l_r - c_r^{A}, 0) = \sum_{p \in P} k_p - |A_p|$. Setting $u_r$ to  $\max(l_r - c_r^{A}, 0)$ for all $r$ ensures that every remaining choice of reviewer decreases $\sum_{r \in R} \max(l_r - c_r^{A}, 0)$ by exactly $1$ for each of the remaining paper choices. If the algorithm terminates with a complete allocation, there will be $\sum_{p \in P} k_p - |A_p|$ more choices, so $\sum_{r \in R} \max(l_r - c_r^{A}, 0)$ will be $0$.

   Finally, we must prove that the allocation remains EF1. The proof from Theorem~\ref{thm:RRR_ef1} applies in this case as well. In that proof, we consider two agents ${p'}$ and $p$. Then we show that if we assume ${p'}$ envies $p$ more than one item after some round $t$ (where each round begins with an assignment to ${p'}$), we can derive a contradiction. We still have that ${p'}$ must prefer the reviewer $p$ got in $t$, $r_p$, to the reviewer ${p'}$ got in $t$, $r_{p'}$. It is now possible that we performed the restriction of remaining reviewers between the assignment of $r_{p'}$ and the assignment of $r_p$. But because the set of reviewers available after the restriction is a subset of the set of reviewers available before the restriction, we still would have attempted to assign $r_p$ to ${p'}$ in $t$ or earlier. The algorithm would therefore have checked for the EF1 violation when assigning $r_p$ to $p$, and we can derive the same contradiction as in the proof of Theorem~\ref{thm:RRR_ef1}.
\end{proof}

\begin{restatable}{theorem}{wrrrrevmin}
    \label{thm:wrrr_revmin}
    \Cref{alg:pickingreviewerslowerbounds}, implemented using the $\PapCrit$ and $\RevCrit$ for $\WRP$, terminates with a WEF1 allocation where all papers receive at most $k_p$ distinct reviewers, all constraints $C$ are satisfied, and no reviewer $r$ is assigned to more than $u_r$ papers. If the algorithm assigns exactly $k_p$ reviewers to each paper, then all reviewers will be assigned to at least $l_r$ papers. 
\end{restatable}
The proof of \Cref{thm:wrrr_revmin} is similar to that of \Cref{thm:WRRR_wef1,thm:rrr_revmin}  and is in \Cref{appx:missing proofs}.

\section{Optimizing Orders for Picking Sequences}
\label{sec:max_usw_RRR}

We have shown how to provably obtain (W)EF1 allocations of reviewers to papers, but have not offered any welfare guarantees so far. We first state that maximizing welfare under $\RRR$ or $\WRP$ is NP-hard, which we prove following the techniques of \citeA{aziz2015possible} and \citeA{aziz2016welfare}. 

\citeA{aziz2016welfare} present the decision problem \possibleutilitarianwelfare: given a fair allocation instance with $n$ agents, $m$ goods, additive valuations functions $v_i$ for all agents $i$, a class of picking sequence mechanisms $\mathcal{C}$, and an integer $t$, is it possible to run a picking sequence in $\mathcal{C}$ and obtain welfare $t$? To obtain hardness results, they use a problem top-$k$ \possibleset \cite{aziz2015possible}: given a fair allocation instance with $n$ agents, $m$ goods, additive valuations functions $v_i$ for all agents $i$, a class of picking sequence mechanisms $\mathcal{C}$, an agent $i$, and an integer $k$, is it possible to run a picking sequence in $\mathcal{C}$ such that $i$ receives its top-$k$ goods?

\begin{proposition}
    \label{prop:rrr_npcomplete}
    Maximizing welfare subject to round-robin (and $\RRR$) is NP-hard.
\end{proposition}

\begin{proof}

\citeA{aziz2015possible} show that for $k \geq 3$, top-$k$ \possibleset is NP-complete for round-robin orders (they refer to round-robin orders as \emph{strict alternation policies}). We use that fact to show that \possibleutilitarianwelfare is NP-complete over the set of round-robin orders.
Given an instance of top-$k$ \possibleset with $k \geq 3$ over round-robin orderings, we construct an instance of \possibleutilitarianwelfare. One agent has utility of $mk^2$ for its $k$ most preferred items, and $0$ for the rest. The other agents have utility at most $k$ for all items. Top-$k$ \possibleset returns true if and only if $mk^3$ utility is achievable.

Reducing an instance of \possibleutilitarianwelfare with welfare threshold $t$ under round-robin to maximizing welfare under $\RRR$ is simple. We can construct an instance where all reviewers review at most one paper. If the round-robin would run for $k$ rounds, then we require each paper to have $k$ reviewers. The additional envy checks in $\RRR$ are only required when some paper cannot select their preferred reviewer even though it has capacity. Since each reviewer cannot be selected more than once total and there are no conflicts of interest, we never invoke that check, and $\RRR$ becomes equivalent to standard round-robin. Therefore, the maximum welfare under this instance of $\RRR$ is at least $t$ if and only if the original \possibleutilitarianwelfare problem evaluates to true. 
\end{proof}
A similar proof extends to show that optimizing welfare over the tie-breaking order $\cal O$ is NP-hard for $\WRP$. As the proof is nearly identical to that of \Cref{prop:rrr_npcomplete}, we relegate it to \Cref{appx:missing proofs}.
\begin{restatable}{proposition}{wrrrnpc}
    \label{prop:wrrrnpc}
      Maximizing welfare subject to weighted picking (and $\WRP$) is NP-hard.
\end{restatable}

In the remainder of  this section, we present a simple greedy approach to approximately maximize the USW of our picking sequence by optimizing over the \emph{ordering} of the papers. 
We present results using $\RRR$ (Algorithm~\ref{alg:rrr}), but all results apply equally well to \Cref{alg:wrrr,alg:pickingreviewers,alg:pickingreviewerslowerbounds}. 
\Cref{thm:main_result} can be used to show that greedy selection of an order $\cal O$ is approximately welfare maximizing for any algorithm implementing \Cref{alg:pickingreviewerslowerbounds}. In particular, when applied to $\WRP$, this implies that greedily selecting a \emph{tie-breaking} order for weighted picking sequences is approximately welfare optimal. Tie-breaking in weighted picking is quite powerful; for example, the tie-breaking order directly determines the first $n$ picks, since all agents start with $0$ items.

We define a function $\USW_{\RRR}(\mathcal{O}, k, R, \{u_r\}_{r \in R}, \{v_p\}_{p \in P})$, which represents the USW from running \RRR~on agents in the order $\mathcal{O}$ with reviewers $R$, reviewer upper limits $u_r$, affinity functions $v_p$, and paper bundle size limit $k$. When it is clear from context, we will drop most of the arguments, writing $\USW_{\RRR}(\mathcal{O})$ to indicate that we run \RRR~with the order $\mathcal{O}$ and all other parameters defined by the current problem instance. Our algorithm, which we call Greedy Reviewer Round Robin (\GRRR), maintains an order $\mathcal{O}$, always adding the paper $p$ which maximizes $\USW_{\RRR}(\mathcal{O} + p)$. 

\begin{algorithm}[t]
    \caption{Greedy Reviewer Round Robin ($\GRRR$)}
    \begin{algorithmic}[1]
        \REQUIRE {$m$ reviewers $R$, limits $\{u_r\}_{r \in R}$, $n$ papers $P$, affinity functions $\{v_p\}_{p \in P}$, bundle size limit $k$}
        \STATE $\mathcal{O} \gets \cal O_\emptyset$
        \FOR{$t \in \{1, \dots , n\}$}
            \STATE $\mathcal{O} \gets \mathcal{O} + p$ where $p$ maximizes $\USW_{\RRR}(\mathcal{O} + p, k, R, \{u_r\}_{r \in R}, \{v_p\}_{p \in P})$ over all $p \in P \setminus \mathcal{O}$
        \ENDFOR
        \RETURN $\mathcal{O}$
    \end{algorithmic}
    \label{alg:greedy_rrr}
  \end{algorithm}
The pseudocode of $\GRRR$ is presented as Algorithm~\ref{alg:greedy_rrr}. It returns an order on papers, which can be directly input to \RRR~to obtain an EF1 allocation of reviewers. 
This algorithm is very simple and flexible. 
It admits trivial parallelization, as the function $\USW_{\RRR}$ can be independently computed for each paper. 
One can also reduce runtime by subsampling the remaining papers at each step. Subsampling weakens the approximation guarantee in theory; while we do not attempt to analyze the approximation ratio of the subsampling approach in this work, we run our largest experiments with this variant, and still obtain high-welfare allocations. 
Let us now establish the welfare guarantees of \GRRR.

We first review important concepts and terms used in the proof. 
A \emph{matroid} \cite{oxley2011matroid} is a pair $(E, \mathcal{I})$ with ground set $E$ and independent sets $\mathcal{I}$, which must satisfy $\emptyset \in \mathcal I$. 
Independent sets must satisfy the inclusion property: $\forall A \subseteq B \in \mathcal{I}$, $A \in \mathcal{I}$, and the exchange property: $\forall A, B \in \mathcal{I}$ with $|A| < |B|$, $\exists e \in B \setminus A$ such that $A \cup\{ e\} \in \mathcal{I}$. 
A \emph{partition matroid} is defined using categories $B_1, B_2, \dots B_b$ such that $B_i \cap B_j = \emptyset$ for all $i, j$ and $\bigcup_{1 \leq i \leq b}B_i = E$, and capacities $d_1, d_2, \dots d_b$; 
the independent sets are $\mathcal{I} = \{I \subseteq E: \forall i, |I \cap B_i| \leq d_i \}$. 
Given two matroids over the same ground set $(E, \mathcal{I}_1)$ and $(E, \mathcal{I}_2)$, the intersection of the two matroids is the pair $(E, \{I : I \in (\mathcal{I}_1 \cap \mathcal{I}_2)\})$. The intersection of two matroids may not be a matroid \cite{oxley2011matroid}.

We also use the notion of a submodular set function; submodular functions formalize the notion of diminishing marginal gains. For a set function $f: 2^E \to \mathbb{R}$, a set $X \subseteq E$, and an element $e \in (E \setminus X)$, we can write the marginal gain of adding $e$ to $X$ under $f$ as $\rho^f_e(X) = f(X+e) - f(X)$ or simply $\rho_e(X)$ if $f$ is understood from context.
Given a set $E$, a function $f: 2^E \to \mathbb{R}$ is \emph{submodular} if for all $X \subseteq Y \subseteq E$ and $e \in E \setminus Y$, $\rho^f_e(X) \geq \rho^f_e(Y)$. 
 A set function is \emph{monotonically non-decreasing} if for all $X \subseteq Y \subseteq E$, $f(X) \leq f(Y)$. We define the notion of $\gamma$-weak submodularity for monotonically non-decreasing, non-negative functions. Given a monotonically non-decreasing, non-negative function $f: 2^E \to \mathbb{R}_{\geq 0}$, we say that $f$ is $\gamma$-weakly submodular if for all $X \subseteq Y \subseteq E$ and $e \in E \setminus Y$, $\gamma\rho^f_e(X) \geq \rho^f_e(Y)$. 
 When $\gamma=1$ we recover submodularity, and we always have $\gamma \geq 1$. 

We show that \GRRR~is equivalent to greedily maximizing a $\gamma$-weakly submodular function over the intersection of two partition matroids.
Consider tuples of the form $(p, i)$ where $p$ is a paper and $i$ represents a position in an order. 
We define a mapping from sets of tuples to orders. 
Consider the set $E = \{(p, i) : p \in P, i \in \{1,\dots,n\}\}$. Define two partition matroids $(E, \mathcal{I}_1)$ and $(E, \mathcal{I}_2)$, such that $\mathcal{I}_1$ forbids duplicating papers, and $\mathcal{I}_2$ forbids duplicating positions. Define $\mathcal{I}_1$ using a category for each paper $p$, where $B^1_p = \{(p, i) : i \in \{1, \dots, n\}\}$, and $\mathcal{I}_1 = \{I \subseteq E : \forall p, |I \cap B^1_p| \leq 1\}$. Likewise, $\mathcal{I}_2$ is defined using a category for each position $i$, where $B^2_i = \{(p, i) : p \in P\}$, and $\mathcal{I}_2 = \{I \subseteq E : \forall i, |I \cap B^2_i| \leq 1\}$.
Any set $Q$ in the intersection of these two matroids can be converted into a paper order $\mathcal{O}_Q$ by sorting $Q$ on the position elements and outputting the paper elements in that order. 
Formally, given any set $Q \in (\mathcal{I}_1 \cap \mathcal{I}_2)$, we construct an order $\mathcal{O}_Q = (S_Q, o_Q)$ by taking $S_Q = \{p \in P: \exists i, (p,i) \in Q\}$. For all $(p, i) \in Q$, let $i' = |\{(p', j) \in Q : j \leq i\}|$ and set $o_Q(p) = i'$. 
An example of this process is given in Table~\ref{tab:rr_ordering_ex}.
We extend this mapping to all subsets of $E$ by sorting on the position elements as a primary key and paper elements as a secondary key, then deleting all but the first tuple for each paper.



\begin{table*}
    \caption{Example showing how sets $Q \subseteq E$ map to orders, and the resulting allocations from executing round-robin. Each paper receives $k=2$ reviewers, and the reviewer upper bounds $u_r$ are $\{2, 1, 2, 1\}$ respectively. The scores (in order) for each reviewer for paper $1$ are $\{2, 5, 2, 7\}$, for paper $2$ are $\{1, 2, 0, 9\}$, and for paper $3$ are $\{4,3,6,3\}$. $Q \subseteq E$ is the set of tuples which map to the order $\cal O_Q$, and $\RRR(\cal O_Q)$ is the allocation resulting from executing $\RRR$ with order $\cal O_Q$. The greedy choices of $\GRRR$ are indicated with asterisks. $\GRRR$ finishes with an EF1 allocation with welfare $21$, although there is an ordering which achieves $27$. Order $[2, 1, 3]$ results in an incomplete allocation.}
        \begin{tabu} {XXXr}
            $Q$ & $\cal O_Q$ & $\RRR(\cal O_Q)$ & $\USW_{\RRR}(\cal O_Q)$ \\
            \midrule
            $\emptyset$ & $[]$ & $\{\}, \{\}, \{\}$ & 0 \\
            $^*\{(1,1)\}$ & $[1]$ & $\{r_4, r_2\}, \{\}, \{\}$ & $12$ \\
            $\{(2,1)\}$ & $[2]$ & $\{\}, \{r_4, r_2\}, \{\}$ & $11$ \\
            $\{(3,1)\}$ & $[3]$ & $\{\}, \{\}, \{r_3, r_1\}$ & $10$ \\

            $\{(1,1), (2, 2)\}$ & $[1, 2]$ & $\{r_4, r_1\}, \{r_2, r_1\}, \{\}$ & $12$ \\
            
            $^*\{(1,1), (3, 2)\}$ & $[1, 3]$ & $\{r_4, r_2\}, \{\}, \{r_3, r_1\}$ & $22$ \\

            $^*\{(1,1), (3, 2), (2, 3)\}$ & $[1, 3, 2]$ & $\{r_4, r_1\}, \{r_2, r_3\}, \{r_3, r_1\}$ & $21$ \\

            $\{(3,1), (2, 2), (1, 3)\}$ & $[3, 2, 1]$ & $\{r_2, r_3\}, \{r_4, r_1\}, \{r_3, r_1\}$ & $27$ \\

            $\{(2,1), (1, 2), (3, 3)\}$ & $[2, 1, 3]$ & $\{r_2, r_1\}, \{r_4, r_1\}, \{r_3\}$ & $23$ \\
        \end{tabu}

        \label{tab:rr_ordering_ex}
\end{table*}
    
With these constructions defined, we observe that maximizing the USW for \RRR~over a fixed number of rounds $k$ is equivalent to the problem $\max_{Q \in (\mathcal{I}_1 \cap \mathcal{I}_2) : |Q|=n} \USW_{\RRR}(\cal O_Q)$ for the matroids defined above. We will show that \GRRR~greedily maximizes a monotonically non-decreasing version of our function over our two partition matroids. 
Next, we show that when our function is $\gamma$-weakly submodular, we can provide a $\gamma$-dependent approximation ratio.

To make $\USW_{\RRR}(\cal O_Q)$ monotonically non-decreasing, we will multiply by a factor of $|Q|^{\alpha}$, where $\alpha$ is defined as the smallest positive number such that $f(Q) = \USW_{\RRR}(\cal O_Q)|Q|^\alpha$ is monotonically non-decreasing. We first prove that \GRRR~greedily maximizes $f(Q)$. Formally, \Cref{lem:gRRR_implicit_monotone} states that \GRRR~selects the element $p$ maximizing $f(Q+(p, i))$ at each iteration.

\begin{lemma}
    \label{lem:gRRR_implicit_monotone}
    Let $f(Q) = \USW_{\RRR}(\cal O_Q)|Q|^\alpha$ for some $\alpha$ such that $f$ is monotonically non-decreasing. Suppose \GRRR~selects paper $p_t$ at each round $t$, resulting in a set of tuples $Q_t$. Then for all $t$, $(p_t, t)$ maximizes $f(Q_{t-1} + (p, i))$ over all $(p, i)$ such that $Q_{t-1} + (p, i) \in (\cal I_1 \cap \cal I_2)$.
\end{lemma}

\begin{proof}
    We first show that any greedy maximizer for $\USW_{\RRR}$ is also a greedy maximizer of $f$. Suppose that $\USW_{\RRR}(\cal O_{Q + (p, i)}) \geq \USW_{\RRR}(\cal O_{Q + (p', i')})$.
    Because $|Q + (p, i)|^\alpha = (|Q| + 1)^\alpha = |Q + (p', i')|^\alpha$, we have
    \begin{align}
    \USW_{\RRR}(\cal O_{Q + (p, i)})|Q + (p, i)|^\alpha \geq \USW_{\RRR}(\cal O_{Q + (p', i')})|Q + (p', i')|^\alpha, \notag
    \end{align}
    as desired.

    We also must prove that we can always simply append to the end of the current ordering (rather than perhaps selecting an arbitrary tuple $(p, i)$). Formally, we want to show that at any point in the algorithm, there is a tuple $(p, |\cal O|)$ that maximizes $\USW_{\RRR}(\cal O + (p, i))$. This is shown via a strong induction argument. For the base case, if some tuple $(p, i)$ maximizes $\USW_{\RRR}(\cal O_\emptyset + (p, i))$, then it is easy to see that $\cal O_\emptyset + (p, i) = \cal O_\emptyset + (p, 1)$ and so we can use $(p, 1)$ without loss of generality. Inductively, assume that we have a set $Q = \{(p_1, 1), (p_2, 2), \dots (p_{|Q|}, |Q|)\}$ and some tuple $(p, i)$ maximizes $f(Q + (p, i))$ such that $Q + (p, i) \in (\cal I_1 \cap \cal I_2)$. 
    Necessarily, $i > |Q|$, since all other positions $i' \leq |Q|$ have been filled in $Q$. Therefore, for any available position $k$, we have that $\cal O_{Q + (p, k)} = [p_1, p_2, \dots p_{|Q|}, p]$ and thus $f(Q + (p, k))$ is the same for all allowed $k$. So without loss of generality, we can assume that we can select a paper for the next available position (as is done in $\GRRR$). 
\end{proof}

The greedy algorithm for maximizing $f(Q)$ terminates when $|Q|=n$, so we must also ensure \GRRR~terminates with an order on all $n$ papers. Although \GRRR~only considers $\USW_{\RRR}(\cal O)$, which may not be monotonically increasing, by construction it runs until reaching a full order over all papers. Thus \GRRR~is equivalent to greedily maximizing $f$.

We are now ready to prove the $1+\gamma$ approximation ratio for \GRRR~(Theorem~\ref{thm:main_result}). Our proof is inspired by the proof in~\cite{fisher1978analysis} that a similar greedy algorithm gives a $\frac{1}{p+1}$-approximation for maximizing a monotone submodular function over the intersection of $p$ matroids. 
However, the introduction of $\gamma$-weak submodularity changes the proof, and we can simplify elements of the proof for our setting. 

\begin{theorem}
    \label{thm:main_result}
    Suppose that $f$ is the monotonically non-decreasing, $\gamma$-weakly submodular function $f(Q) = \USW_{\RRR}(\cal O_Q)|Q|^\alpha$. The set $Q^{\alg}$ returned by \GRRR~satisfies $f(Q^{\alg}) \geq \frac{1}{1+\gamma}f(Q^*)$, where $\cal O_{Q^*}$ is the optimal paper order for \RRR. Because $|Q^{\alg}| = |Q^*|$, this implies that $Q$ approximates the maximum welfare with constant $\frac{1}{1+\gamma}$.

\end{theorem}

\begin{proof}
    Let $Q_t$ represent the subset of $Q^{\alg}$ after step $t$ of \GRRR, 
    where we add the element $(p_t, t)$ to $Q_{t-1}$. 
    Let $(p_t^*, t)$ denote the pair in $Q^*$ which places paper $p_t^*$ in position $t$. 
    Denote $L = |Q^* \setminus Q^{\alg}|$. Consider the elements of $Q^* \setminus Q^{\alg} = \{(p^\ast_{t_1}, t_1), \dots (p^\ast_{t_L}, t_L)\}$, ordered so that $t_1 < t_2 < \dots t_L$. 
    Let $Q^{\alg} \cup \{(p^\ast_{t_1}, t_1), \dots (p^\ast_{t_l}, t_l)\}$ be denoted as $Q^{\alg}_l$ (with $Q^{\alg}_0 = Q^{\alg}$). By monotonicity of $f$, $f(Q^\ast)$ is bounded from above by:
    \begin{align}
        f(Q^{\alg} \cup Q^*) = f(Q^{\alg}) + \sum\limits_{l=1}^L \rho_{(p^\ast_{t_l}, t_l)}(Q^{\alg}_{l-1}). \label{ineq1}
    \end{align}
    By $\gamma$-weak submodularity of $f$, we have that 
    \begin{align}
        \rho_{(p^\ast_{t_l}, t_l)}(Q^{\alg}_{l-1})
        \leq  \gamma \rho_{(p^\ast_{t_l}, t_l)}(Q_{t_l-1}). \label{ineq2}
    \end{align}
    Equality~\ref{ineq1} and inequality~\ref{ineq2} imply that
    \begin{align}
        f(Q^\ast) \leq f(Q^{\alg}) + \gamma \sum\limits_{l=1}^L \rho_{(p^\ast_{t_l}, t_l)}(Q_{t_l-1}),  \notag
    \end{align}
    which (again by monotonicity of $f$) is bounded by
    \begin{align}
        f(Q^{\alg}) +\gamma \sum\limits_{t=1}^n \rho_{(p^\ast_{t}, t)}(Q_{t-1}). 
        \label{ineq4}
    \end{align}
    Next, we claim that for all $t$,
    \begin{align}
        \rho_{(p^\ast_{t}, t)}(Q_{t-1}) \leq \rho_{(p_{t}, t)}(Q_{t-1}). \label{greedy_best}
    \end{align}
    At step $t$, the greedy algorithm chose to add $(p_t, t)$ to $Q_{t-1}$, with $p_t$ maximizing $f(Q_{t-1} + (p_t,t))$. If $(p_t^\ast, j)$ is not present in $Q_{t-1}$ for any $j$, then the greedy algorithm would have considered adding $(p_t^\ast, t)$ and determined that $p_t$ was better. Suppose that $(p_t^\ast, j) \in Q_{t-1}$ for some $j$. The greedy algorithm proceeds by filling positions from left to right, so $j \le t-1$. By the definition of our mapping from sets to orders, $p_t^\ast$ will take position $j$ and ignore  $(p_t^\ast, t)$. 
    Thus $\rho_{(p_t^\ast, t)}(Q_{t-1}) = 0 \leq \rho_{(p_t, t)}(Q_{t-1})$. In either case, inequality~\eqref{greedy_best} holds.
    Combining~\eqref{ineq4} with~\eqref{greedy_best} yields
    \begin{align}
        f(Q^\ast) &\leq f(Q^{\alg}) + \gamma \sum\limits_{t=1}^n \rho_{(p_{t}, t)}(Q_{t-1}) = (1+\gamma)f(Q^{\alg}). \notag
    \end{align}
\end{proof}

When $\gamma=1$ (and thus $f$ is submodular), Theorem~\ref{thm:main_result} yields a $\frac{1}{2}$-approximation guarantee, which beats the $\frac{1}{3}$-approximation guarantee provided by~\cite{fisher1978analysis}. The greedy algorithm is a tight $\frac{1}{2}$-approximation for submodular maximization in the \emph{unconstrained} regime~\cite{buchbinder2012tight}, which our result matches even though we operate in a constrained (albeit less general) space.

\section{\FairSeq: a Fast and Fair Reviewer Assignment Algorithm}
\label{sec:fairsequence}

We have shown extensively how to obtain fair and approximately welfare efficient reviewer assignments. However, $\GRRR$ can be prohibitively slow. 

\begin{proposition}
\label{prop:grrr_runtime}
The runtime of $\GRRR$ is $O(kmn^4)$, where $k$ is the number of reviewers assigned per paper, $m$ is the total number of reviewers, and $n$ is the total number of papers.
\end{proposition}

\begin{proof}
    There are $n$ positions to fill in the round-robin order. At each position, we have to check at most $n$ papers to determine which is the greedy choice. For each choice of paper, we run $\RRR$ on at most $n$ papers over at most $k$ rounds. When a paper gets assigned a reviewer during $\RRR$, we may have to attempt to assign at most $m$ reviewers, and each time we attempt to assign, we have to check against at most $n$ other papers for EF1 violations. Thus, each individual assignment during $\RRR$ takes $O(nm)$ time, and there are $O(nk)$ iterations of this. $\RRR$ takes $O(kmn^2)$ time, and we have to run it $O(n)$ times to select the greedy maximizer for each position, over $n$ positions. Thus, total runtime is $O(kmn^4)$.
\end{proof}

In initial experiments we found that $\GRRR$ can finish in a reasonable amount of time (one or two days) on smaller conferences ($m, n < 1000$). Even this runtime is not ideal, as conference organizers often must determine reviewer assignments over the course of several days and typically try multiple assignments using different formulas for affinity scores.  In this section, we describe an approach that sacrifices theoretical  guarantees for speed and improved empirical welfare. Our algorithm, $\FairSeq$, uses the weighted picking sequence described by \citeA{wef1}, similarly to $\WRP$ described in Section~\ref{sec:wef1}. However, rather than breaking ties using an order that is fixed ahead of time, we will break ties greedily \emph{for each tie-break}. Thus, rather than having to run the picking sequence multiple times to determine the greedy choice in the order, we can run a single picking sequence with the determination of the greedy choice consisting of a simple maximum operation over a matrix. Because \FairSeq uses the picking sequence derived from \citeA{wef1}, it still maintains the WEF1 criterion. $\FairSeq$ is described in \Cref{alg:fairsequence}. $\FairSeq$ has a provably faster runtime than $\GRRR$ (\Cref{prop:fairseq_runtime}), and we show that in practice it is significantly faster than all existing reviewer assignment algorithms (\Cref{sec:experiments}).

\begin{proposition}
\label{prop:fairseq_runtime}
The runtime of $\FairSeq$ is $O(kmn^3)$, where $k$ is the maximum number of reviewers assigned per paper, $m$ is the total number of reviewers, and $n$ is the total number of papers.
\end{proposition}

\begin{proof}
    $\FairSeq$ fills each of the $nk$ positions in the picking sequence sequentially. At each position, we select from at most $n$ papers, and each paper selects from $m$ reviewers. Finally, for each possible reviewer-paper pair, we may have to check $n$ other papers to avoid WEF1 violations. 
\end{proof}
In many conference settings, $\FairSeq$ drastically improves on the $O(kmn^3)$ upper bound.  In contrast to all algorithms presented up to this point, once we find a single valid reviewer-paper pair to assign, we can automatically rule out any reviewer-paper pairs with lower affinity score. We can also apply another shortcut; when checking that the WEF1 criterion is met on line $6$, it is sufficient to only compare against the papers $p'$ that have $v_{p'}(r) > v_{p'}(r')$ for some $r' \in A_{p'}$. When $v_{p'}(r) \leq v_{p'}(r')$ for all $r \in A_p$ and $r' \in A_{p'}$, $\frac{v_{p'}(A_p)}{|A_p|} \leq \frac{v_{p'}(A_{p'})}{|A_{p'}|}$. This is equivalent to WEF1 when $|A_p| = k_p$ and $|A_{p'}| = k_{p'}$, as is the case in a complete allocation.   In large conferences with many subject areas, this can rule out most comparisons across major subject areas.

$\FairSeq$ maintains a WEF1 guarantee as long as it terminates with a complete allocation.

\begin{proposition}
    
    \label{thm:fairsequence_fair}
    If $\FairSeq$ returns a complete allocation, that allocation is WEF1.
\end{proposition}

\begin{proof}
    This holds by the final condition on line $6$ of \Cref{alg:fairsequence}. This condition implies that after every assignment of some reviewer $r$ to some paper $p$, we have checked for all other papers ${p'}$ if WEF1 is met. 
\end{proof}

Despite its benefits, $\FairSeq$ may not always terminate with a complete allocation.\footnote{In fact, OpenReview reported a large AI conference on which $\FairSeq$ fails to return a complete allocation. The details of this conference are confidential, and OpenReview were not authorized to share the data with us.} Therefore, we introduce a second algorithm $\FairSeqUnchecked$ that will always terminate with a complete allocation, as long as the constraints $C$ only represent conflict of interest constraints. Although the second algorithm is not guaranteed to be WEF1, the fact that the algorithm is based on a weighted picking sequence implies that the allocation will still be roughly fair. $\FairSeqUnchecked$ operates similarly to $\FairSeq$. However, we no longer perform the additional checks used to ensure $\FairSeq$ is WEF1. In addition, because the picking sequence assigns reviewers irrevocably, when a conference is highly constrained it may be possible for some paper to have no feasible assignments later in the process. When this occurs, we have to revoke (and replace) some of the earlier assignments in order to free up a feasible reviewer to assign. We use a heuristic approach, where we run the algorithm with multiple different $\beta$ values, indicating the amount by which we allow welfare to drop during any of these swaps. In order to allow $p$ to swap $r'$ for $r$, we require that $v_p(r) \geq \beta v_p(r')$. Setting $\beta = 1$ requires that no swaps can reduce the welfare of $p$, and provides very little flexibility in assignments. Setting $\beta = 0$ allows any swap, potentially impacting welfare and fairness quite a bit, but ensuring termination with a complete and constraint-satisfying allocation. Although these swaps might again cause WEF1 violations, by progressively decreasing $\beta$ we can limit the amount of welfare lost during this process. $\FairSeqUnchecked$ is presented in \Cref{alg:fairsequenceunchecked}.


        

\begin{algorithm}[t]
    \caption{$\FairSeq$}
    \begin{algorithmic}[1]
        \REQUIRE {Reviewers $R$, papers $P$, reviewer upper limits $u_r$, reviewer lower limits $l_r$, affinity functions $v_p$, bundle size limits $k_p$, constraints $C$, bound modification function $h$}

        \STATE Initialize allocation $A$ as $A_p \gets \emptyset$ for all papers $p \in P$
        
        \WHILE{$\exists p: |A_p| < k_p$}
            \STATE $\{k_p\}_{p \in P}, \{l_r\}_{r \in R}, \{u_r\}_{r \in R} \gets h(A, \{k_p\}_{p \in P}, \{l_r\}_{r \in R}, \{u_r\}_{r \in R})$

            \STATE $P^\ast \gets \argmin_{p}\frac{|A_p|}{k_p}$

            \FOR{$p \in P^\ast$}
                \STATE Select $r_p \in R$ as the reviewer maximizing $v_p(r_p)$ such that $c_r^{A} < u_r$, $r \notin A_p$, $\neg C(p, r, A)$, and all $p' \neq p$ 
                satisfy WEF1 with respect to $A_p + r$
            \ENDFOR
            
            \IF{No valid pair $p, r_p$ exists}
                \RETURN $A$
            \ELSE
                \STATE Select $(p^\ast, r_{p^\ast})$ that maximizes $v_{p^\ast}(r_{p^\ast})$
                \STATE $A_{p^\ast} \gets A_{p^\ast} + r_{p^\ast}$
            \ENDIF
        \ENDWHILE
        \RETURN $A$
    \end{algorithmic}
    \label{alg:fairsequence}
  \end{algorithm}

\begin{algorithm}[t]
    \caption{$\FairSeqUnchecked$}
    \begin{algorithmic}[1]
        \REQUIRE {Reviewers $R$, papers $P$, reviewer upper limits $u_r$, reviewer lower limits $l_r$, affinity functions $v_p$, bundle size limits $k_p$, conflicts of interest $C$, sorted beta values $\vec \beta$, bound modification function $h$}

        \STATE $v_{\texttt{max}} \gets \max_{p, r} v_p(r)$
        \FOR{$\beta \in \vec \beta$}

            \STATE Initialize allocation $A$ as $A_p \gets \emptyset$ for all papers $p \in P$
            
            \WHILE{$\exists p: |A_p| < k_p$}
                \STATE $\{k_p\}_{p \in P}, \{l_r\}_{r \in R}, \{u_r\}_{r \in R} \gets h(A, \{k_p\}_{p \in P}, \{l_r\}_{r \in R}, \{u_r\}_{r \in R})$

               \STATE $P^\ast \gets \argmin_{p}\frac{|A_p|}{k_p}$

                \FOR{$p \in P^\ast$}
                    \STATE Select $r_p \in R$ as the reviewer maximizing $v_p(r_p)$ such that $c_r^{A} < u_r$, $r \notin A_p$, $\neg C(p, r, A)$
                \ENDFOR
                
                \IF{No valid pair $p, r_p$ exists}
                    \STATE $R_{avail} \gets {r \in R: c_r^{A} < u_r}$
                    \STATE $V \gets \{(r, -1) : r \in R_{avail}\}$
                    \STATE $V \gets V + \{(r, p) : p \in P, r \in A_p\}$
                    \STATE $E \gets \{\left((r', p'), (r, p), v_{\texttt{max}} + v_p(r)-v_p(r')\right): r' \notin A_p, \neg C(p, r'), v_p(r') \geq \beta v_p(r)\}$
                    \STATE Set graph $G \gets (V, E)$
                    \STATE $Z = {(r_0, -1), (r_1, p_1), \dots (r_{z}, p_{z})} \gets$ shortest weighted path in $G$ from $R_{avail}$ to a paper $p_{z}$ with reviewer $r_{z} \in A_{p_{z}}$ s.t.  $\exists p^\ast \in P^\ast$ with  $r_{z} \notin A_{p^\ast}$ and $\neg C(p^\ast,r_{z})$
                    \STATE Swap reviewers along $Z$ ($A_{p_i} \gets A_{p_i} - r_i + r_{i-1}$ for $i \in \{1, \dots z\}$
                    \STATE $p^\ast = \argmax_{\{p \in P^\ast: r_{z} \notin A_p, \neg C(p, r_{z})\}} v_p(r_{z})$
                    \STATE $r_{p^\ast} \gets r_{z}$
                    \IF{No valid $p^\ast$ exists}
                        \STATE Restart with lower value of $\beta$
                    \ENDIF
                \ELSE
                    \STATE Select $(p^\ast, r_{p^\ast})$ that   maximizes $v_{p^\ast}(r_{p^\ast})$
                \ENDIF
                \STATE $A_{p^\ast} \gets A_{p^\ast} + r_{p^\ast}$
            \ENDWHILE
            \RETURN $A$
        \ENDFOR
    \end{algorithmic}
    \label{alg:fairsequenceunchecked}
  \end{algorithm}

\begin{theorem}
    \label{thm:fairsequenceunchecked}
    If the constraints $C$ consist of only conflict of interest constraints, then $\FairSeqUnchecked$ returns a complete allocation satisfying all assignment constraints, if such an allocation exists. 
\end{theorem}

\begin{proof}
It suffices to consider the case when $\beta=0$. Suppose the picking sequence reaches a point where there are no valid assignments possible to the papers $P^* = \argmin_p \frac{|A_p|}{k_p}$, but some reviewers have not yet reached their maximum review load. Denote the partial allocation up to this point as $A$. Consider any paper $p \in P^*$. There is a complete allocation $A^{c}$ satisfying all assignment constraints, where $A^{c}_p = \{r_1, r_2, \dots r_{k_p}\}$. $|A_p| < |A^c_p|$, so there is some reviewer $r \in A^{c}_p \setminus A_p$. If $r$ has not reached its load upper bound in A, we could add $r$ to $A_p$, since $C$ contains only conflict of interest constraints and $r$ does not have a conflict of interest with $p$. So $r$ must have reached its load upper bound under $A$. Consider any paper that has been assigned $r$ in $A$. One such paper, $p'$, must have $r \notin A^{c}_{p'}$, and instead there is some $r' \notin A_{p'}$ with $r' \in A^{c}_{p'}$. Again, if $r'$ is available, we can simply assign $r'$ to ${p'}$ and $r$ to $p$. If not, the argument repeats for $r'$ -- there must be some paper currently assigned $r'$ that does not receive $r'$ in $A^{c}$.  This sequence must eventually terminate, since there are a finite number of reviewers and papers. Thus, eventually we will find a path to the set of reviewers with remaining review load, and we will be able to make transfers along this path to our paper $p$, allowing us to increase the total number of assignments by $1$ and move on with the rest of the picking sequence.
\end{proof}

$\FairSeqUnchecked$ may have a very long runtime (\Cref{prop:fairsequ_runtime}, with proof in \Cref{appx:missing proofs}). However, $\FairSeq$ often terminates with a complete allocation in practice, making $\FairSeqUnchecked$ unnecessary. In addition, we expect that the picking sequence will run almost to completion before needing to make any of the swaps proscribed by $\FairSeqUnchecked$. This means that the additional runtime of finding and executing these swaps will typically not have a strong impact on the overall runtime. 

\begin{restatable}{proposition}{fairsequruntime}
\label{prop:fairsequ_runtime}
The runtime of $\FairSeqUnchecked$ is $O(|\vec \beta| kn(nm+m)^2)$, where $|\vec \beta|$ is the number of values of $\beta$ in $\vec \beta$, $k$ is the maximum number of reviewers assigned per paper, $m$ is the total number of reviewers, and $n$ is the total number of papers.
\end{restatable}

\section{Empirical Fairness, Efficiency, and Runtime Analysis}
\label{sec:experiments}

In this section, we compare $\GRRR$ and $\FairSeq$ against baselines on three conference datasets. We find $\FairSeq$ is over an order of magnitude faster than all baselines and much fairer (in terms of WEF1 and Gini) than all baselines except PeerReview4All (\PRFA) \cite{stelmakh2019peerreview4all}.

\subsection{Experimental Design}

We run experiments on three conference datasets: Medical Imaging with Deep Learning (MIDL), Conference on Computer Vision and Pattern Recognition (CVPR), and the 2018 iteration of CVPR\footnote{According to \cite{shah2022challenges}, the ``CVPR'' dataset is from 2017, though the paper introducing the dataset  does not list the year \cite{kobren2019paper}.}. 

MIDL, CVPR, and CVPR'18 are standard datasets in the reviewer assignment literature, provided as pre-computed affinity score matrices, reviewer load upper bounds, and paper demands. MIDL is an order of magnitude smaller than CVPR and CVPR'18, and CVPR'18 is slightly less challenging than CVPR due to a higher ratio of reviewer availability to paper demand. 

A summary of the data statistics appears in Table \ref{table:data-summary}. For CVPR'18, while affinities are between 0 and 11, most are between 0 and 1. In addition, reviewer load bounds vary by reviewer but range between 2 and 9.




\begin{table}
    \centering
    \caption{Data summary for the three datasets. Here, $n$ is the number of papers; $m$ is the number of reviewers; val. range is the range of reviewer-paper affinities; $k$ is the bound on the number of reviewers per paper; $u_r$ is the upper bound on papers/reviewer. }
    \begin{tabularx}{0.9\linewidth}{lXXXXX}
    \toprule
    \textbf{Name:} & $n$ & $m$ & val. range&$k$ & $u_r$ \\
    \midrule
    \textbf{MIDL}: & $118$  & $177$  &  $[-1,1]$ &  $3$  &    $4$ \\
    \midrule
    \textbf{CVPR}:& $2623$ & $1373$ & $[0,1]$ & $3$ & $6$ \\
    \textbf{CVPR'18}: & $5062$ & $2840$ & $[0,11]$ & $3$ & $2-9$ \\
    
    \bottomrule
    \end{tabularx}
    \label{table:data-summary}
\end{table}

We compare our methods to the \FairFlow algorithm~\cite{kobren2019paper}, the Toronto Paper Matching System (\TPMS)~\cite{charlin2013toronto}, and PeerReview4All (\PRFA)~\cite{stelmakh2019peerreview4all}. \FairFlow is currently implemented in OpenReview, and it is in widespread use. \TPMS (also in widespread use) provides an upper bound on welfare without fairness guarantees. \PRFA was used by ICML 2020~\cite{stelmakh2021towards}. 
All algorithms are publicly available on Github.\footnote{$\FairSeq$ and \FairFlow: \url{https://github.com/openreview/openreview-matcher}, $\GRRR$, \TPMS, and Constrained Round Robin: \url{https://github.com/justinpayan/ReviewerAssignmentCode}, \PRFA: \url{https://github.com/niharshah/peerreview4all}.}

Following \citeA{kobren2019paper} and \citeA{stelmakh2019peerreview4all}, we only run one iteration of \PRFA on CVPR and CVPR'18. 
On those two conferences, \PRFA maximizes the minimum paper score, but stops before maximizing the next smallest score. 

We also implemented the Constrained Round Robin algorithm \cite{aziz2019constrained}.
CRR is approximately 40 times slower than GRRR on MIDL, taking 400 seconds instead of 10. GRRR takes about 18 hours to run on CVPR. Extrapolating these results, we can expect CRR to require a month of computation time or longer on CVPR (it did not terminate in our experiments). 
Given its infeasible runtime, we did not continue to compare against CRR as a baseline. 
Due to the size of the CVPR'18 dataset, we subsample 100 papers at each iteration of $\GRRR$ rather than testing every available paper. We recorded means and standard deviations over $5$ runs, but found little variation in solution quality from run to run.

\subsection{Fairness and Efficiency}

Fairness and efficiency results for all conferences are included in Table~\ref{tab:results}. 

We report the USW, minimum paper score, and number of EF1 violations for each algorithm. For each setting with at least one violation of the EF1 criterion, we report the total number of papers that envy some other paper more than $1$ reviewer, and the total number of papers that are envied by some paper by more than $1$ reviewer. We report the USW as the percentage of the optimal value (given by \TPMS). 
For an allocation $A$, the number of EF1 violations is the number of ordered pairs of papers $p \neq p'$ failing EF1. There are $n^2 - n$ total potential violations.

\begin{table*}
    \caption{High-level statistics for all conferences. We include a breakdown of EF1 violations where applicable, with the total number of papers that have envy more than one reviewer and the total number of papers that are envied by other papers up to more than one reviewer.}
    \begin{tabu} to \textwidth {lXX[1.5,r]X[1.5,r]X[1.5,r]X[1.5,r]X[1.5,r]}
        \toprule
         & Alg. & USW (\% OPT) & Min Score & EF1 Viol. & Num. Envious & Num. Envied \\
        \midrule
            & \FairFlow & 100\% & 0.94 & 0 & -- & -- \\
       & \TPMS & 100\% & 0.90 & 0 & -- & -- \\
         MIDL\hspace{1.5em}    & \PRFA & 98\% & 0.92 & 0 & -- & -- \\
            & $\GRRR$ & 98\% & 0.83 & 0 & -- & -- \\
            & $\FairSeqShort$ & 99\% & 0.87 & 0 & -- & -- \\
            
        \midrule
            & \FairFlow & 96\% & 0.77 & 23244 & 688 & 1058 \\
             & \TPMS & 100\% & 0.00 & 471256 & 717 & 2097 \\
         CVPR   & \PRFA\hspace{2em} & 94\% & 0.77 & 83& 14 & 75 \\
            & $\GRRR$ & 88\% & 0.00 & 0& -- & -- \\
            & $\FairSeqShort$ & 92\% & 0.00 & 0 &-- & -- \\

        \midrule
            & \FairFlow & 97\% & 9.79 & 23 & 21 & 23 \\
             & \TPMS & 100\% & 1.37 & 134 & 65 & 108 \\
         CVPR'18   & \PRFA\hspace{2em} & 97\% & 12.68 & 2 & 1 & 2 \\
            & $\GRRR$ & 94\% & 1.78 & 0 & -- & -- \\
            & $\FairSeqShort$ & 96\% & 1.74 & 0 & -- & -- \\

    \end{tabu}
    \label{tab:results}
\end{table*}

\FairFlow and \TPMS have very high levels of EF1 violations on CVPR. 
Although some EF1 violations may be permissible, a large number of violations implies that many papers received unnecessarily imbalanced assignments relative to other papers.  We also show the number of papers that have envy for another paper, that cannot be rectified by dropping $1$ reviewer. This analysis shows that the EF1 violations are generally spread out across roughly $10\%$ of the papers, but those papers typically envy a much larger number of papers. In other words, there is a small proportion of papers that received an unduly low quality of reviewer assignments, and would have strongly preferred many other papers' assignments.

\begin{figure}
    \centering
    \begin{subfigure}{0.45\textwidth}
    \centering
        \vspace{-2.4cm}
        \includegraphics[width =\textwidth]{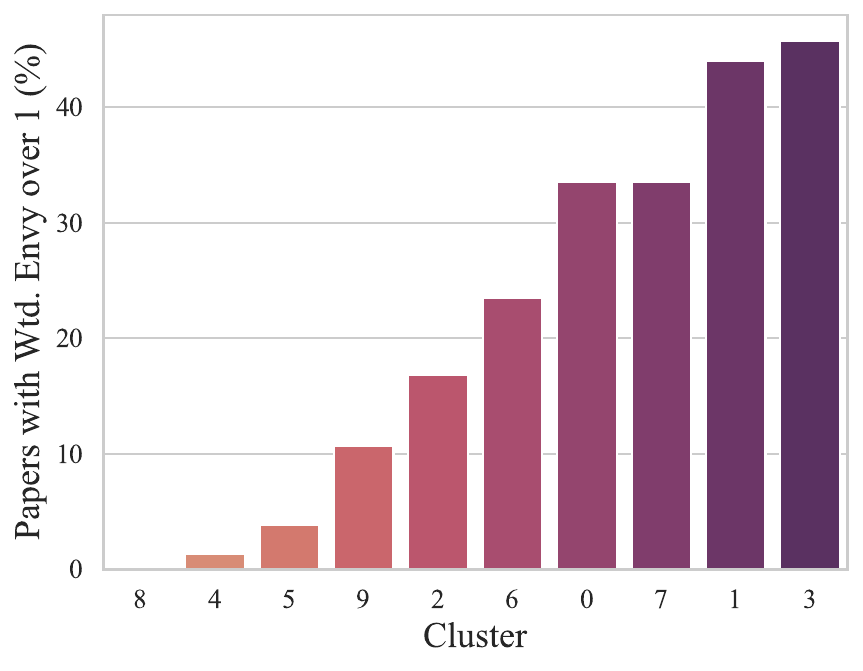}   
        \caption{Number of papers with envy over $1$ reviewer for any other paper, by paper cluster.}
           \label{fig:qualrevsbyarea}
    \end{subfigure}
    \begin{subfigure}{0.45\textwidth}
    \vspace{.05cm}
    \centering
        \includegraphics[width=\textwidth]{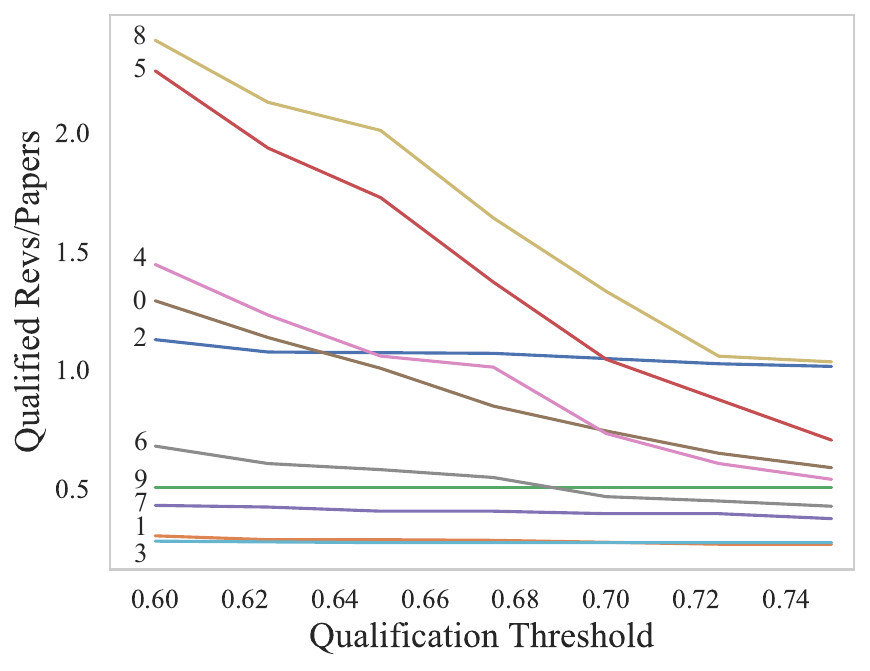}
        \caption{The number of qualified reviewers per cluster, divided by the cluster size. We determine if a reviewer is qualified for a subject area by taking the average of the top $4$ affinity scores for that reviewer over papers in the subject area. We compare this average against a qualification threshold (shown on x-axis). }
        \label{fig:qualrevs}
    \end{subfigure}

    \caption{
        Envy violations and reviewer quality in CVPR, using \TPMS assignment. The paper clusters with the most EF1 violations have fewer qualified reviewers relative to the cluster size. 
    }
 
\end{figure}

To further understand the potential sources and impacts of EF1 violations, we analyze the distribution of papers that have envy over one reviewer for any other paper in the \TPMS assignment for CVPR. Although the paper identities are not available in the dataset, we can simulate a subject area classification by clustering papers. We represent each paper $p$ as the vector  of affinity scores between $p$ and all reviewers, $\vec v_p \in \R^m$. We then cluster these vectors into $10$ clusters using agglomerative clustering under a Euclidean metric and Ward linkage function.\footnote{\url{https://scikit-learn.org/stable/modules/generated/sklearn.cluster.AgglomerativeClustering.html#sklearn.cluster.AgglomerativeClustering}} \Cref{fig:qualrevsbyarea} shows the percentage of papers in each cluster that envy at least one other paper more than $1$ reviewer. We can see that some of the clusters have very high percentages of strongly envious papers (up to $40\%$), while others have almost no envy.

What drives these WEF1 violations? In \Cref{fig:qualrevs}, we plot the number of ``highly qualified'' reviewers for a subject area, divided by the number of papers in the area. We define a reviewer as highly qualified for a subject if the average affinity over the top four most similar papers in the subject exceeds some threshold. \Cref{fig:qualrevs} demonstrates the number of ``highly qualified'' reviewers per subject area as we vary the qualification threshold.
The subject areas with the highest fraction of submissions violating EF1 also tend to have lower ratios of qualified reviewers to papers, indicating that EF1 violations accumulate in more heavily resource-constrained subject areas. 
It is possible that these subject areas suffer from a dearth of qualified reviewers; thus, any papers that receive qualified reviewers are inevitably envied by other papers in that subject area.

\begin{table*}
        \caption{Inequality statistics for $\GRRR$, $\FairSeq$, \FairFlow, and \PRFA. We compute the bottom decile and quartile of papers by score, then report the mean and standard deviation for both low-percentile blocks. We also report the Gini coefficient of all paper scores and the sum of the envy across all paper pairs, where lower is better for both.}
        \begin{tabu} to \textwidth {lXX[2,r]X[2,r]X[r]X[2,r]}            \toprule
             & Alg. & Lowest 10\% & Lowest 25\% & Gini &  \hspace{2em}Envy \\
            \midrule
            \multirow{4}*{MIDL}    & \FairFlow & $1.051 \pm .072$ & $1.186 \pm .131$ & .146 &  .501 \\
                & \PRFA & $1.069 \pm .082$ & $1.211 \pm .135$ & \textbf{.127} &  \hspace{2em}.448 \\
                & \GRRR & $.995 \pm .095$ & $1.164 \pm .157$ & .145 & \hspace{2em}.834 \\
                & $\FairSeqShort$ & $1.040 \pm .085$ & $1.191 \pm .139$ & .140 & \textbf{.146} \\
                
            \midrule
            \multirow{4}*{CVPR}        & \FairFlow & $.838 \pm .032$ & $.908 \pm .068$ & .233 &  64462 \\
             & \PRFA & $1.065 \pm .150$ & $1.324 \pm .247$ & \textbf{.145} &  \textbf{9287} \\
                   & \GRRR & $.898 \pm .176$ & $1.110 \pm .217$ & .183 &  22400 \\
                   & $\FairSeqShort$ & $.978 \pm .139$ & $1.197 \pm .218$ & .169 & 10602 \\

            \midrule
    
           \multirow{4}*{CVPR'18}     & \FairFlow & $11.053 \pm .536$ & $12.519 \pm 1.805$ & .151 &  6940 \\
                & \PRFA & $15.280 \pm .952$ & $16.668 \pm 1.348$ & \textbf{.103} &  \textbf{2480} \\
                   & \GRRR & $8.923 \pm 2.890$ & $12.220 \pm 3.528$ & .168 &  \hspace{2em}28840 \\
                   & $\FairSeqShort$ & $10.084 \pm 2.540$ & $12.950 \pm 3.182$ & .154 & 17419 \\


        \end{tabu}
        \label{tab:inequality_stats}
    \end{table*}
    
\begin{figure}
\centering
\begin{subfigure}{0.45\textwidth}
    \hspace{-.75cm}
    \centering
    \includegraphics[width=\textwidth]{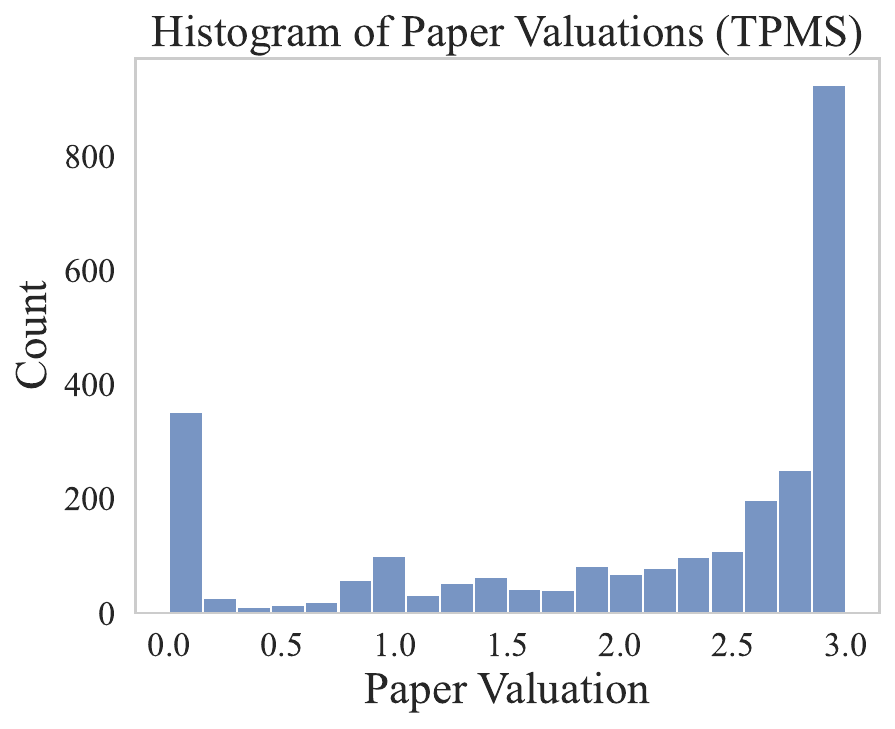}
\end{subfigure}%
\begin{subfigure}{0.45\textwidth}
    \centering
     \includegraphics[width=\textwidth]{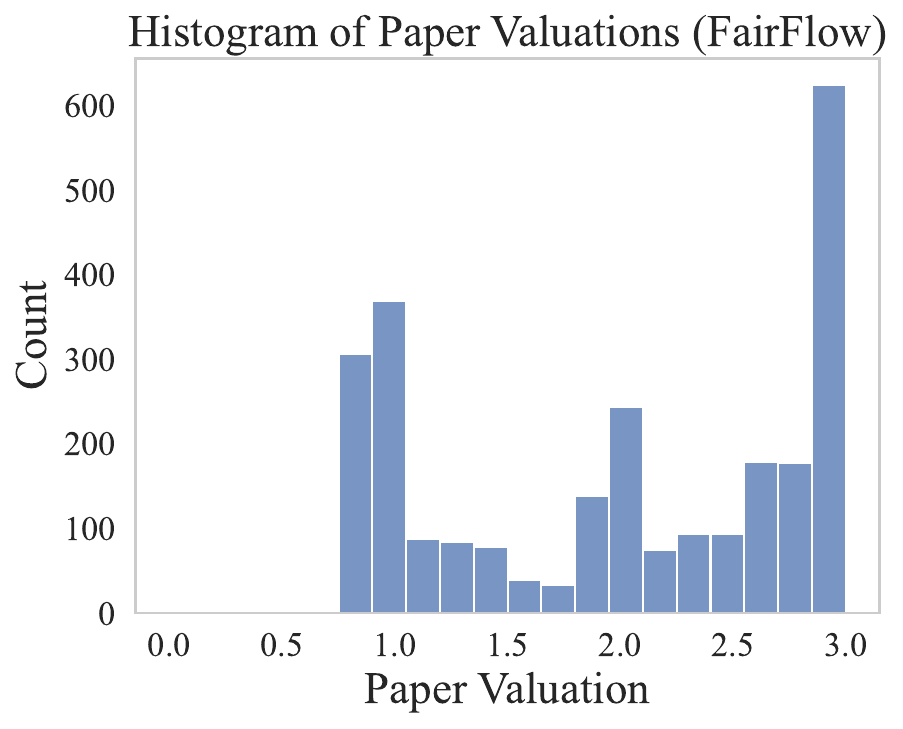}
\end{subfigure}
\begin{subfigure}{0.45\textwidth}
    \centering
     \includegraphics[width=\textwidth]{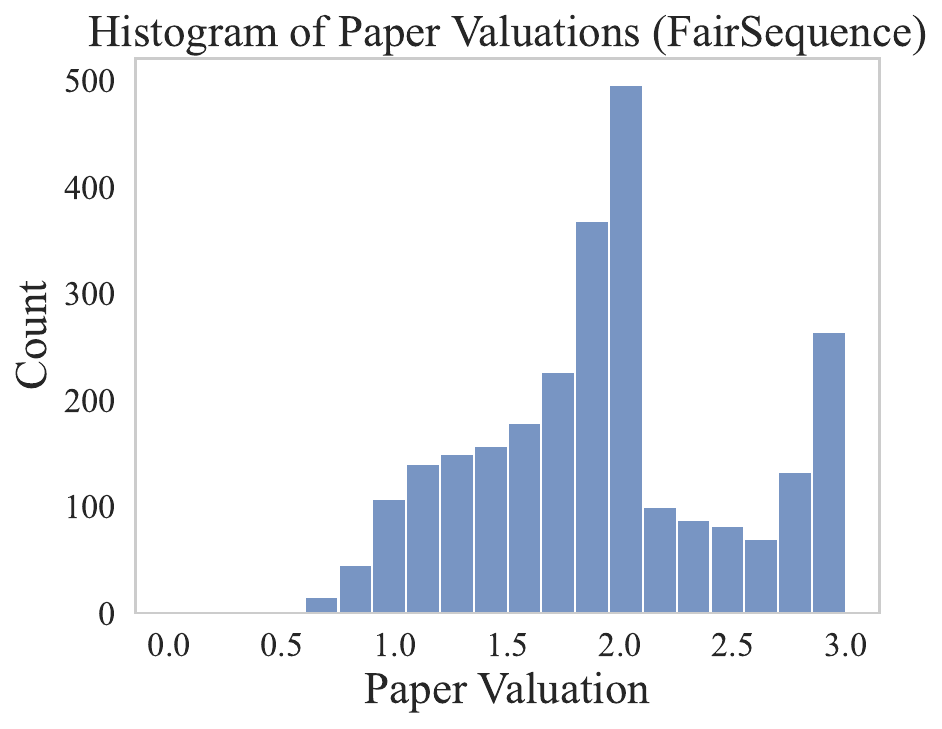}
\end{subfigure}
\begin{subfigure}{0.45\textwidth}
    \centering
     \includegraphics[width=\textwidth]{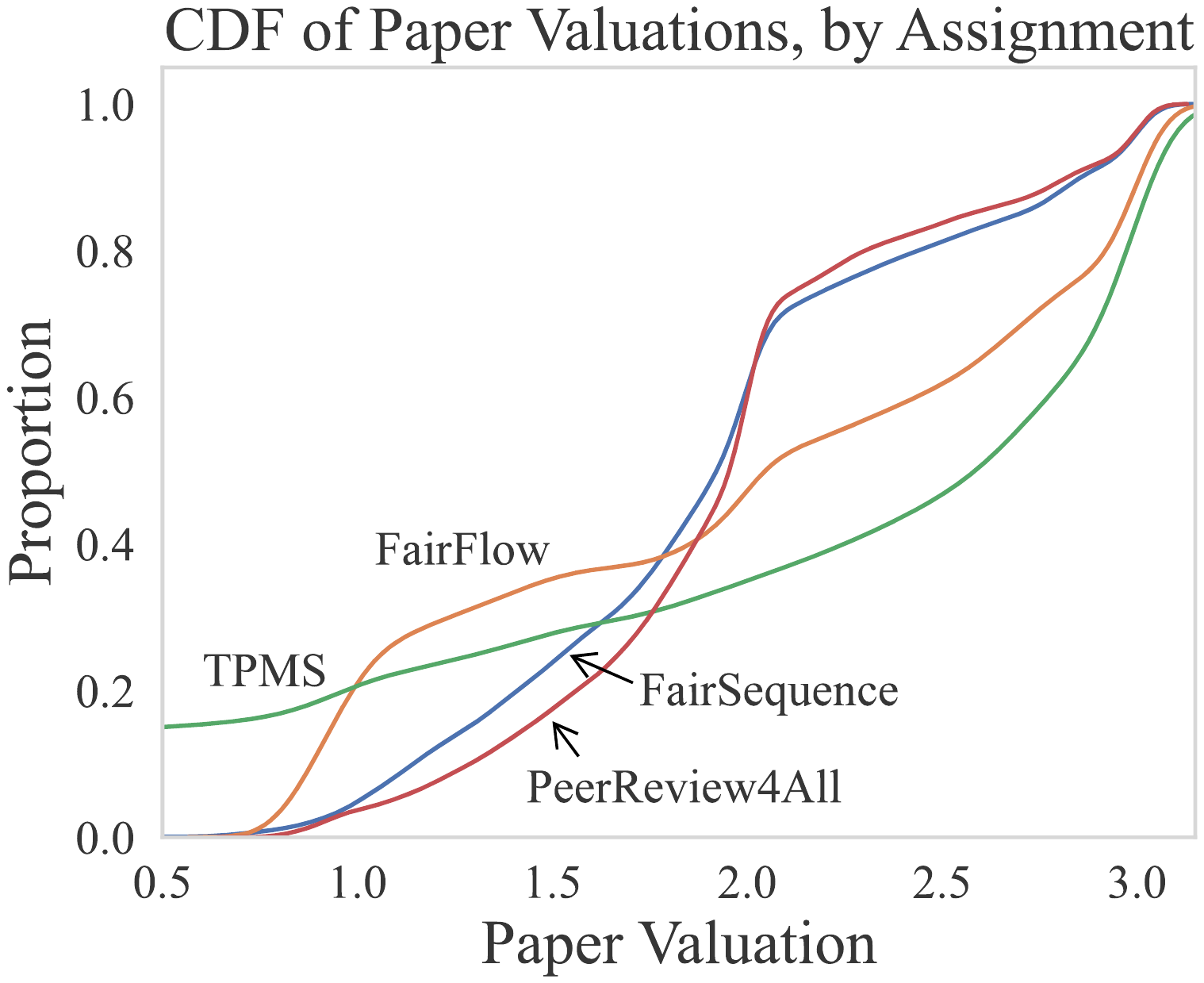}
\end{subfigure}

    \caption{Distribution of paper valuations for CVPR under \TPMS,  \FairFlow, and \FairSeq. \FairFlow, which maximizes the minimum paper score, results in a less fair overall distribution of paper scores than \FairSeq. We also show the cumulative distribution of paper scores for \TPMS, \FairFlow,  \FairSeq, and \PRFA assignments. The bottom decile and quartile for \FairSeq and \PRFA are much higher than the bottom decile and quartile for \FairFlow, and all three improve over \TPMS.}
    \label{fig:FairFlowunfair}
\end{figure}

For additional measures of inequality, we compute the mean and standard deviation of paper scores for the bottom decile and quartile of papers per allocation. 
We consider allocations to be more fair if they allocate higher scores to these disadvantaged papers. 
We also calculate the Gini coefficient for each allocation, a standard measure of inequality \cite{gini1936measure}. A higher Gini coefficient indicates more inequality, so lower is better for this metric. Finally, we report the sum of the total envy over all ordered paper pairs $p$ and $p'$, $\sum_{p,p' \in P} \max \{(v_p(A_{p'}) - v_p(A_p)), 0\}$. The results are summarized in Table~\ref{tab:inequality_stats}. 
\Cref{fig:FairFlowunfair} shows the full distribution of paper scores under \TPMS, \FairFlow, and $\FairSeq$ for CVPR.\footnote{Similar plots for MIDL and CVPR'18 are included in \Cref{appx:plots}. MIDL shows almost no variation across algorithms, while CVPR'18 shows a more nuanced picture. \FairSeq outperforms \FairFlow on low percentiles, but \TPMS shows stronger bottom quartile scores than either. However, \TPMS creates a cluster of very low scoring papers that are fully or partially mitigated by the other approaches.} 
$\FairSeq$ significantly outperforms \FairFlow in the scores given to lower decile and quartile papers, as well as in the Gini index. 
These metrics are closer on MIDL and CVPR'18, which we have already seen are less constrained settings. 
\Cref{fig:FairFlowunfair} explains the reason for the decreased fairness of \FairFlow on CVPR; although it maximizes the minimum paper score, it leaves many papers clustered near the minimum paper score. 
$\FairSeq$ smoothly shifts the entire distribution of paper scores rightward. 
\PRFA is much fairer than both algorithms, but it does not handle variable paper demands, and has a higher computational overhead, as we will see in \Cref{subsec:runtime}.



\subsection{Runtime Analysis}
\label{subsec:runtime}

\begin{figure}
    \centering
\includegraphics[width=.7\textwidth]{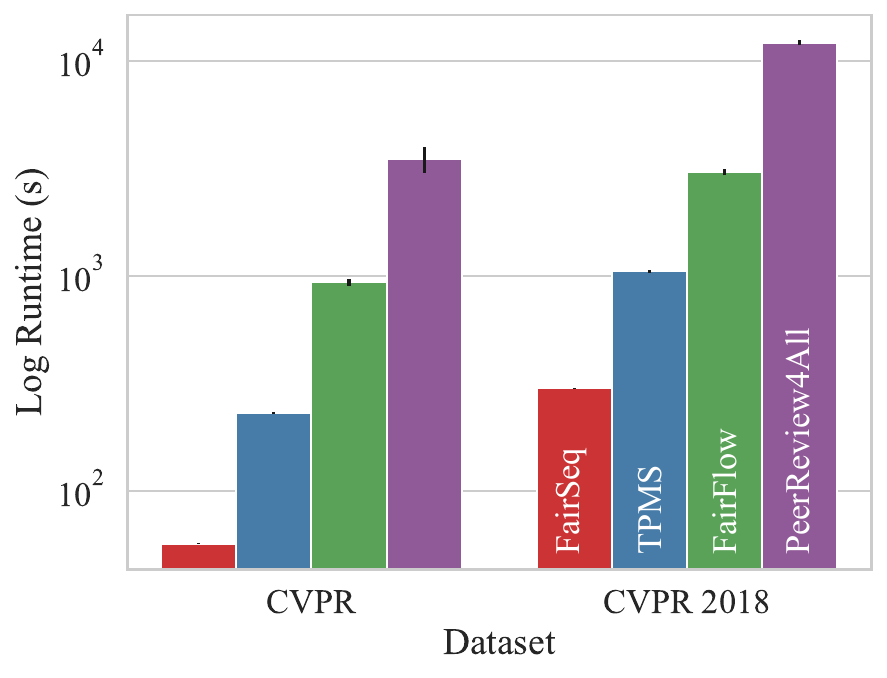}
    \caption{Runtimes of \FairSeq, \TPMS, \FairFlow, and \PRFA on CVPR and CVPR 2018. Runtimes are not reported for MIDL since all algorithms take $<10$ seconds to run. 
    \FairSeq is at least $3$ times faster than \TPMS, the second-fastest competitor.}
    \label{fig:runtimes}
\end{figure}

Perhaps the biggest benefit of \FairSeq is its greatly improved runtime.  We display the runtimes (in seconds) of \PRFA, \FairFlow, \TPMS, and \FairSeq in Figure~\ref{fig:runtimes}.  
$\GRRR$ takes longer than a day to run on CVPR and CVPR'18, so we also do not include it in the analysis. We find that $\FairSeq$ is orders of magnitude faster than even \TPMS, which has no fairness guarantees. Further, $\FairSeq$ has  been implemented in pure Python while the other three rely primarily on highly optimized optimization tools (\PRFA and \TPMS were implemented using Gurobi, and much of the computation in \FairFlow was done using Google's OR-Tools). Despite these implementation differences, $\FairSeq$ is consistently at least $3$ times faster than \TPMS, $10$ times faster than \FairFlow, and $50$ times faster than \PRFA. 

\subsection{Additional Experiments}
We perform some additional experiments, details of which are in the appendix. We estimate $\alpha$ and $\gamma$ for $\GRRR$ on MIDL, CVPR, and CVPR'18 in \Cref{appx:alpha_gamma}, and we discuss the topic of fairness to reviewers in \Cref{appx:rev_loads}.



\section{Conclusion}



Our algorithms $\GRRR$ and $\FairSeq$ ensure (W)EF1. $\FairSeq$ strongly outperforms the two state-of-the-art fair algorithms \FairFlow and \PRFA in runtime, and outperforms \FairFlow on most fairness measures. $\GRRR$ and $\FairSeq$ are easy to implement and understand, and their simple formulations give them the flexibility to handle many additional constraints. 
Speed and flexibility matter a great deal in practice; in fact, these are the main benefits of \FairFlow over an alternative introduced in \citeA{kobren2019paper} with better fairness and welfare. OpenReview regularly uses \FairFlow, but has not implemented the other algorithm to date.

The reviewer assignment problem provides interesting constraints, complicating the standard fair allocation setting. 
We demonstrate that a straightforward round-robin allocation, combined with a novel optimization technique on paper orders, finds EF1 allocations with high USW in the reviewer assignment setting. 
Our approach of optimizing over \emph{orders} for round-robin allocations is of independent interest, and may inspire further study of optimal round-robin allocations. We then extended this approach to incorporate variable paper demands and reviewer lower bounds, before finally presenting our algorithm $\FairSeq$. Empirically, we showed that $\FairSeq$ is fast, fair, flexible, and has high welfare. Our analysis suggests that maximum egalitarian welfare may not be enough for ensuring distributional fairness. In addition, our study shows initial empirical evidence for a connection between the (W)EF1 fairness criterion and the Gini inequality index; further study is warranted in this direction.

Finally, there are many other applications of different fairness, efficiency, robustness, and incentive compatibility constraints from the fair allocation literature to problems in peer review, which could bring some much-needed rigor to this fundamental process.

\acks{Justin Payan and Yair Zick are supported by NSF grant CISE:IIS:RI: 2327057. We thank Harold Rubio, Purujit Goyal, Melisa Bok, and Andrew McCallum from OpenReview for help developing our problem statement and testing initial versions of \FairSeq, Vignesh Viswanathan for teaching us about submodular optimization, and the reviewers of Coop AI at NeurIPS 2021, WINE 2021, AAMAS 2022, GAIW at AAMAS 2022, and IJCAI 2022.  This work was performed in part using high performance computing equipment obtained under a grant from the Collaborative R\&D Fund managed by the Massachusetts Technology Collaborative.}





\vskip 0.2in
\bibliographystyle{theapa}
\bibliography{abb,JAIR/main}

\clearpage

\appendix

\section{Missing Proofs}
\label{appx:missing proofs}

\largerevs*

\begin{proof}

    Algorithm~\ref{alg:rrr} only refuses to assign a reviewer $r$ to a paper $p$ when $r$ is assigned to too many papers, $r$ has already been assigned to $p$, or some other paper (to which we have previously attempted to assign $r$) ``objects'' to the assignment. Thus if we have assigned $l < kn$ distinct reviewers under Algorithm~\ref{alg:rrr}, it must be the case that there is a reviewer $r$ that we have not considered for any paper. Because there are at least $kn$ distinct reviewers, we can see that during any round of the algorithm, there will be such an unconsidered reviewer. Thus in any round, a paper can always be assigned some reviewer that has never been considered for any paper, and the selection will not be refused. This proves that the allocation returned by Algorithm~\ref{alg:rrr} is complete, and we have EF1 from Theorem~\ref{thm:RRR_ef1}.
\end{proof}

\wrrrrevmin*

\begin{proof}
    The distinctness of reviewers per paper, reviewer upper bounds $u_r$, satisfaction of $C$, and upper limit on $k_p$ reviewers per paper are satisfied for the same reasons described in Theorem~\ref{thm:WRRR_wef1}. The lower limits are satisfied for complete assignments for the same reason described in the proof of Theorem~\ref{thm:rrr_revmin}. 

    We must prove the allocation remains WEF1. Again, we consider the moment when a paper $p$ is assigned a reviewer $r$. For all ${p'}$ to which we have attempted to assign $r$, we check that ${p'}$ does not have weighted envy for $p$ over one item if we assign $r$ to $p$. This check will still occur even if the reviewer restriction occurs between ${p'}$'s attempt at $r$ and $p$'s assignment of $r$.  

    When we have $|S_{p'} \cap S_p| = 0$, we have not attempted to assign any reviewers to ${p'}$ that we have attempted to assign to $p$, including $r$. 
    We stated in the proof of Theorem~\ref{thm:WRRR_wef1} that for any reviewer $r_{p'}$ assigned to ${p'}$, $v_{p'}(r_{p'}) > v_{p'}(r_p)$ for any $r_p$ that $p$ received after ${p'}$ received $r_{p'}$. This claim is sufficient for the proof from~\cite{wef1} to go through. If the reviewer restriction occurs before ${p'}$ received $r_{p'}$, clearly we still have $v_{p'}(r_{p'}) > v_{p'}(r_p)$ for any $r_p$ that $p$ received after ${p'}$ received $r_{p'}$ (it is the same argument, but in the restricted problem setting). If the restriction happens after ${p'}$ received $r_{p'}$, we see that ${p'}$ is being assigned reviewers from a superset of reviewers compared to the previous case. Thus we still have $v_{p'}(r_{p'}) > v_{p'}(r_p)$ for any $r_p$ that $p$ received after ${p'}$ received $r_{p'}$.
\end{proof}

\wrrrnpc*

\begin{proof}
    \citeA{aziz2016welfare} show that \possibleutilitarianwelfare is NP-complete for recursively balanced orders (orders where each agent picks once in each round). Given an instance of \possibleutilitarianwelfare for recursively balanced orders with welfare threshold $t$, we can reduce to the problem of maximizing welfare over $\WRP$. If there are $k$ rounds in the recursively balanced picking sequence, then assume there are $k$ rounds in $\WRP$. Again all reviewers review at most one paper. When that is the case, $\WRP$ is exactly equivalent to the weighted picking sequence from \cite{wef1} (that is, no allocations are rejected because any attempted assignment will satisfy WEF1). This means that the sequence of assignments in $\WRP$ could have been achieved by a recursively balanced picking sequence in the original problem, and therefore we achieve a maximum welfare of at least $t$ in $\WRP$ if and only if the original \possibleutilitarianwelfare problem evaluates to true. 
\end{proof}

\fairsequruntime*
\begin{proof}
    Like $\FairSeq$, $\FairSeqUnchecked$ fills each of the $nk$ positions in the picking sequence sequentially. At each position, we may have to visit all nodes and edges of the exchange graph. Since there are $O\left((n+1)m\right)$ nodes, this operation costs $O((nm + m)^2)$ time. In the worst case, we have to run the main loop of the algorithm $\abs{\vec \beta}$ times.
\end{proof}

\section{Additional Plots}
\label{appx:plots}

\begin{figure}
\centering
\begin{subfigure}{0.45\textwidth}
    \hspace{-.75cm}
    \centering
    \includegraphics[width=\textwidth]{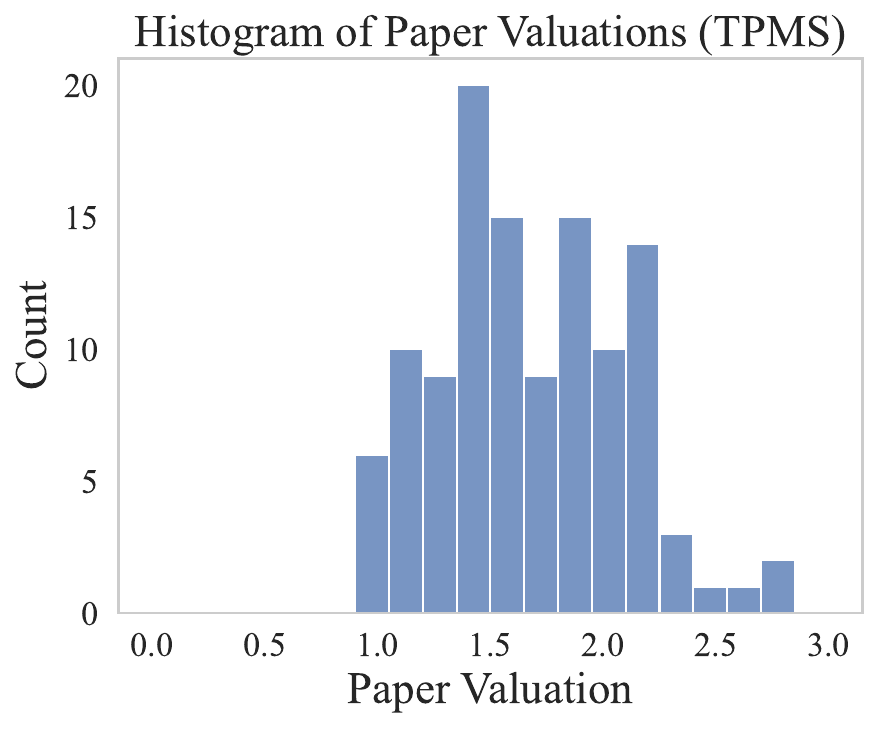}
\end{subfigure}%
\begin{subfigure}{0.45\textwidth}
    \centering
     \includegraphics[width=\textwidth]{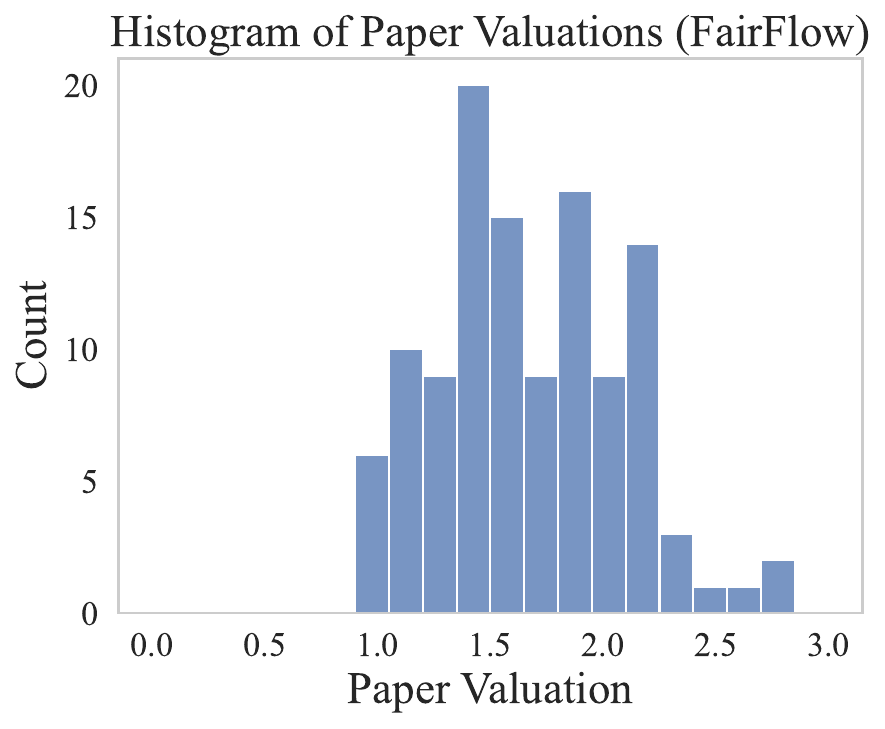}
\end{subfigure}
\begin{subfigure}{0.45\textwidth}
    \centering
     \includegraphics[width=\textwidth]{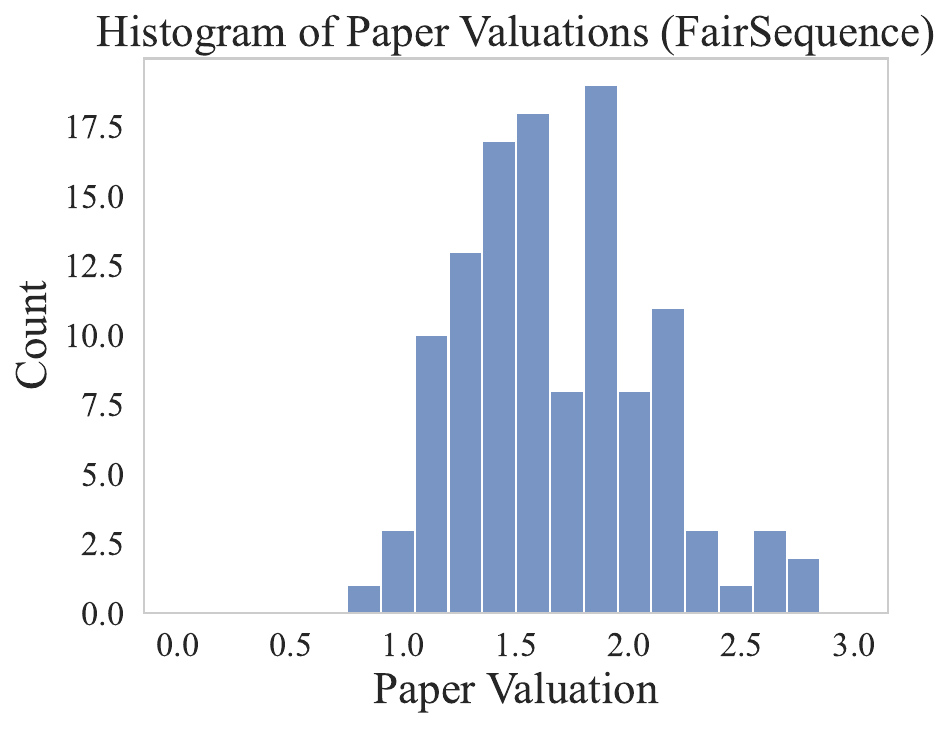}
\end{subfigure}
\begin{subfigure}{0.45\textwidth}
    \centering
     \includegraphics[width=\textwidth]{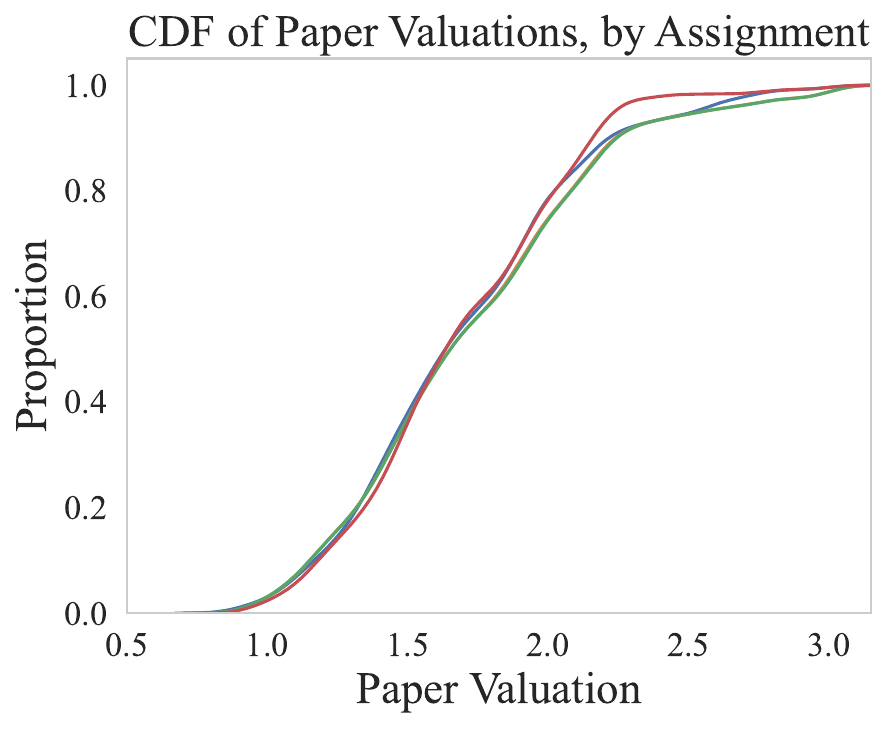}
\end{subfigure}

    \caption{Distribution of paper valuations for MIDL under \TPMS,  \FairFlow, and \FairSeq. We also show the cumulative distribution of paper scores for \TPMS, \FairFlow,  \FairSeq, and \PRFA assignments. }
    \label{fig:midldistrib}
\end{figure}

\begin{figure}
\centering
\begin{subfigure}{0.45\textwidth}
    \hspace{-.75cm}
    \centering
    \includegraphics[width=\textwidth]{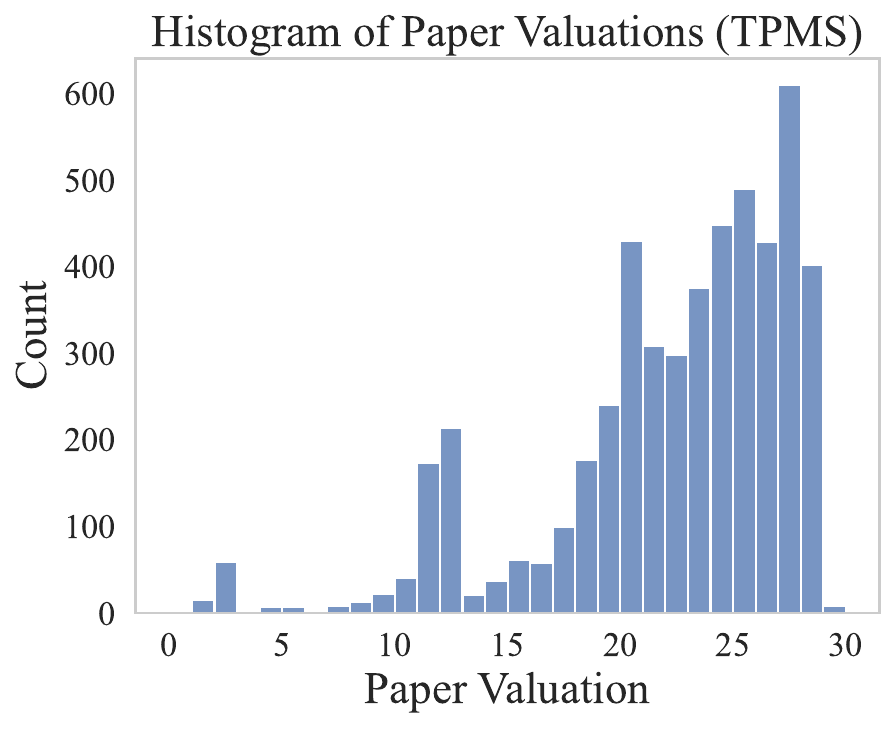}
\end{subfigure}%
\begin{subfigure}{0.45\textwidth}
    \centering
     \includegraphics[width=\textwidth]{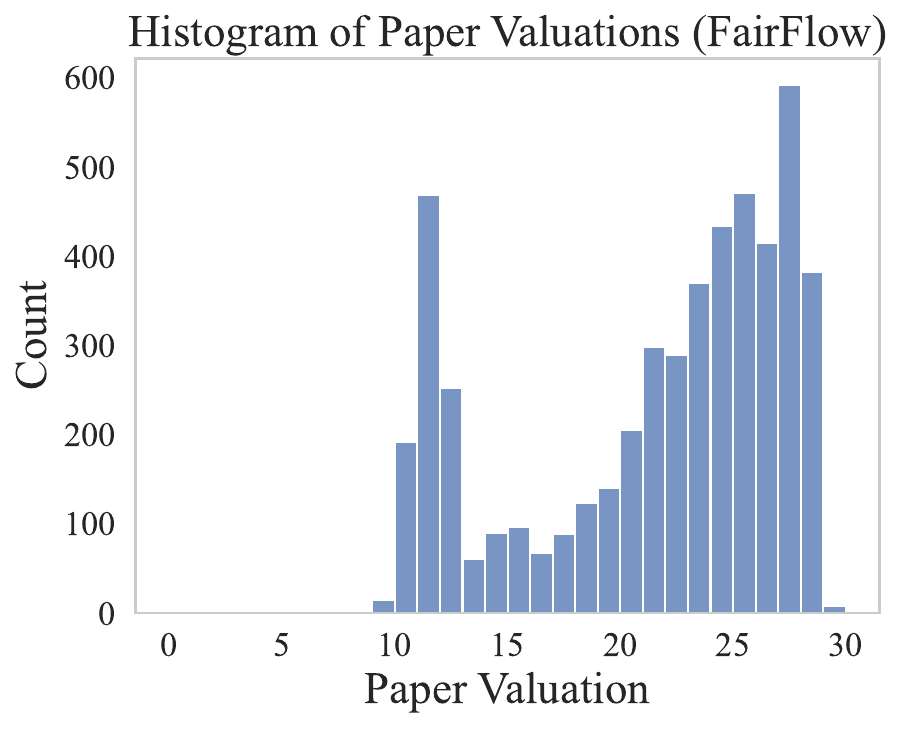}
\end{subfigure}
\begin{subfigure}{0.45\textwidth}
    \centering
     \includegraphics[width=\textwidth]{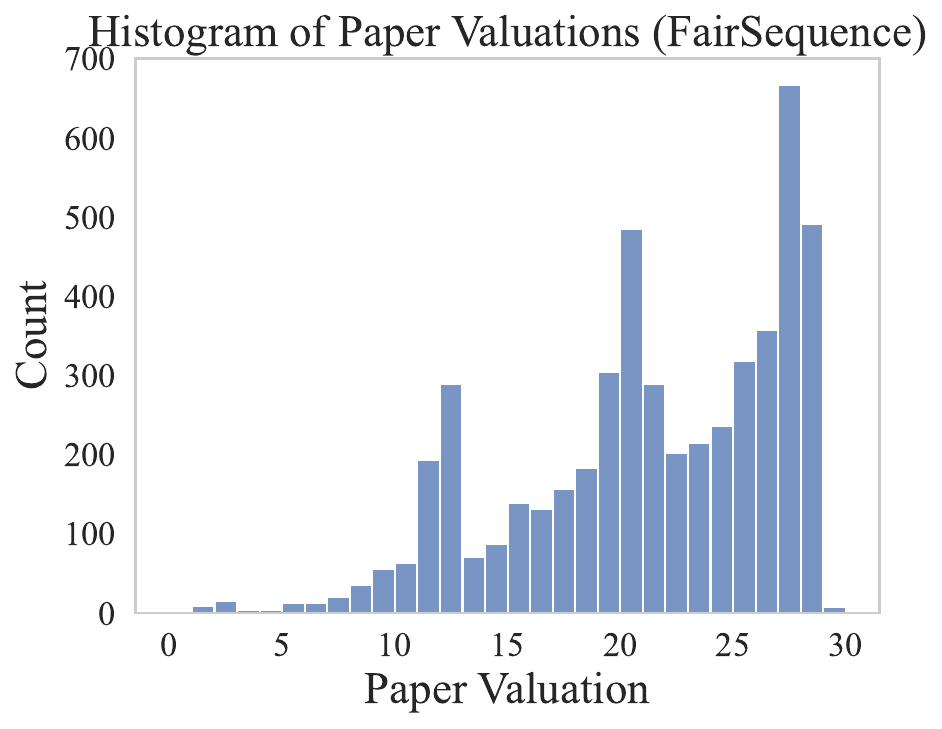}
\end{subfigure}
\begin{subfigure}{0.45\textwidth}
    \centering
     \includegraphics[width=\textwidth]{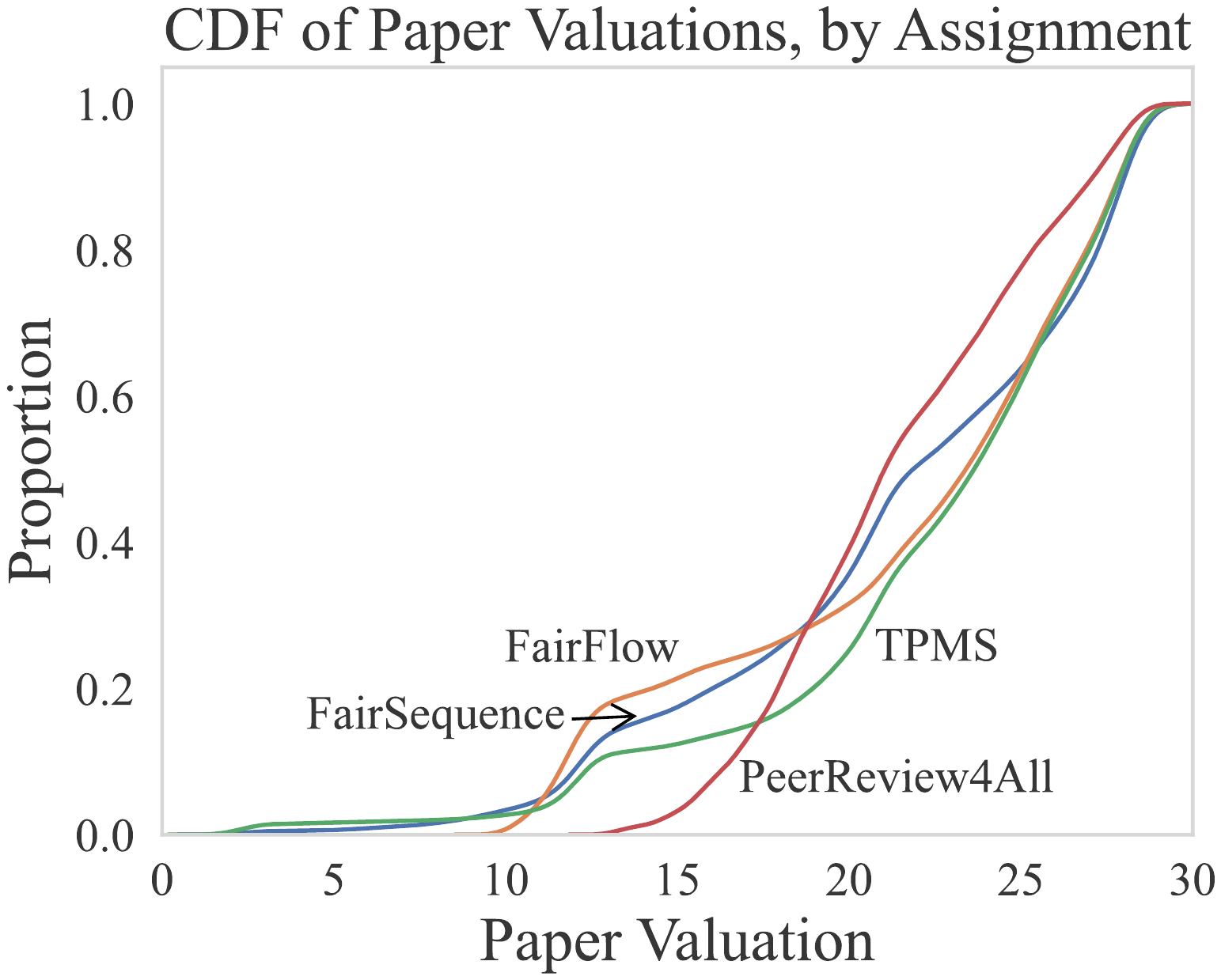}
\end{subfigure}

    \caption{Distribution of paper valuations for CVPR'18 under \TPMS,  \FairFlow, and \FairSeq. We also show the cumulative distribution of paper scores for \TPMS, \FairFlow,  \FairSeq, and \PRFA assignments. }
    \label{fig:cvpr18distrib}
\end{figure}

\Cref{fig:midldistrib,fig:cvpr18distrib} show the full distribution of paper scores on MIDL and CVPR'18 for each algorithm. MIDL shows little variation across assignment algolithms. On CVPR'18, we see that  \TPMS has a surprisingly high bottom quartile paper score. However, the full distributional plots demonstrate a cluster of very low-scoring papers for \TPMS that are largely mitigated by \FairSeq and \FairFlow (at the expense of lower mid-range percentile scores). Once again, \PRFA seems to maintain very strong fairness guarantees relative to all baselines.


\section{Estimation of Empirical Guarantees}
\label{appx:alpha_gamma}

We estimate $\alpha$ and $\gamma$ for $\GRRR$ on MIDL, CVPR, and CVPR'18.
$\gamma$ is rather large for CVPR and CVPR'18. It is possible that other conferences or other application areas would yield welfare functions that are closer to monotonically increasing and submodular, leading to lower values of $\gamma$. 

For any order $\mathcal{O}$ and any paper $p \notin \mathcal{O}$, we must have that
$\USW_{\RRR}(\mathcal{O}+p)|\mathcal{O} + p|^\alpha \geq \USW_{\RRR}(\mathcal{O})|\mathcal{O}|^\alpha$. When $\USW_{\RRR}(\mathcal{O}+p) > \USW_{\RRR}(\mathcal{O})$, any positive $\alpha$ will satisfy this inequality. We estimate $\alpha$ by sampling orders $\mathcal{O}$ and papers $p \notin \mathcal{O}$, and we take our estimate to be slightly greater than the maximum $\alpha$ found for any $\mathcal{O}$ and paper $p$. 
For MIDL, we found that no sampled $\mathcal{O}$ and $p$ violate monotonicity, so we set $\alpha$ to be $0.01$. 
Using our estimated $\alpha$ values, we then estimate $\gamma$. 
Here, we sample $X$ and $Y$ so that $X \subseteq Y$, and $e \notin Y$. We need $\gamma \geq \frac{\rho_e(Y)}{\rho_e(X)}$ for all samples. 
Similarly to our $\alpha$ estimate, we compute $\frac{\rho_e(Y)}{\rho_e(X)}$ for all samples and then estimate $\gamma$ to be slightly greater than the maximum value. 
We found in all experiments that our chosen $\alpha$ parameter led to all positive marginal gains during the $\gamma$ estimation, improving our confidence in the $\alpha$ estimates. 
The results are displayed in Table~\ref{tab:alpha_gamma}.

\begin{table}
 \caption{Estimated $\alpha$ and $\gamma$ parameters for all three conference datasets. Values of $\alpha$ close to $0$ indicate the $\USW_{\RRR}$ function is close to monotonically increasing on that dataset. The approximation ratio for Algorithm~\ref{alg:greedy_rrr} is $1 + \gamma$, so higher $\gamma$ yields a looser approximation guarantee. We did not find any pairs $\mathcal{O}$ and $p \notin \mathcal{O}$ such that $\USW_{\RRR}(\mathcal{O} + p) < \USW_{\RRR}(\mathcal{O})$ for MIDL, so $\alpha$ can be set arbitrarily close to $0$.}
        \begin{tabu} to \columnwidth {XX[r]X[r]}
            & $\alpha$ & $\gamma$ \\
            \midrule
            MIDL & $^*0.01$ & 1.21 \\
            CVPR & 1.03 & 50.62 \\
            CVPR'18 & 0.51 & 17.41 \\
        \end{tabu}

        \label{tab:alpha_gamma}
\end{table}

\section{Fairness to Reviewers}
\label{appx:rev_loads}


\begin{figure}
    \centering
\includegraphics[width=\textwidth]{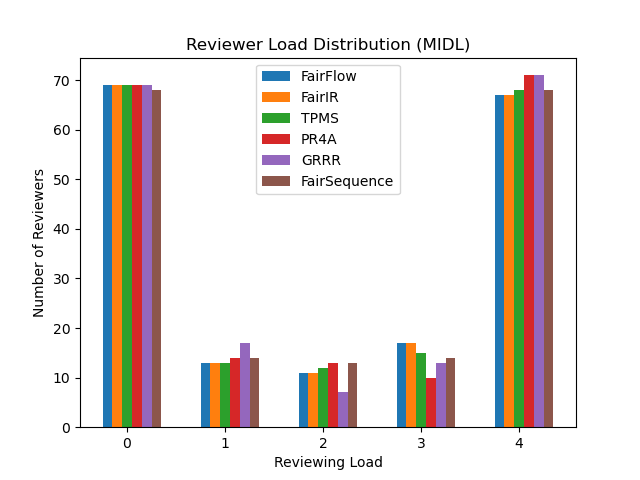} 
      \caption{Distribution of reviewer loads for all algorithms on MIDL.}
    \label{fig:rev_loads_midl}
\end{figure}

\begin{figure}
    \centering
\includegraphics[width=\textwidth]{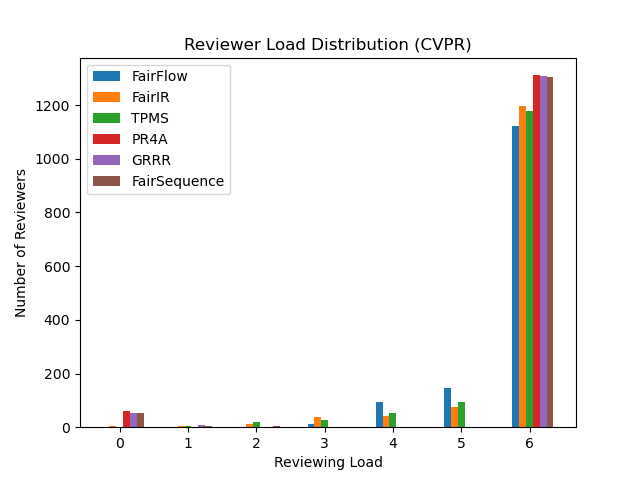} 
      \caption{Distribution of reviewer loads for all algorithms on CVPR.}
    \label{fig:rev_loads_cvpr}
\end{figure}

\begin{figure}
    \centering
\includegraphics[width=\textwidth]{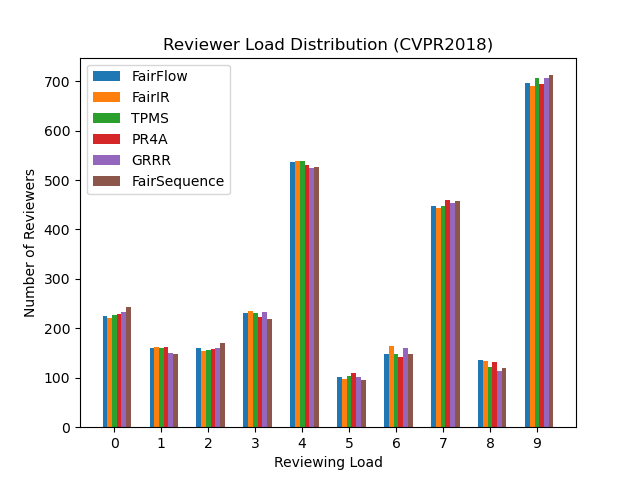} 
      \caption{Distribution of reviewer loads for all algorithms on CVPR'18.}
    \label{fig:rev_loads_cvpr2018}
\end{figure}

We take the position throughout this paper that is more appropriate to treat reviewers, rather than papers, as goods. Paper reviewing is generally viewed as a chore, not a benefit, whereas papers do benefit from appropriate reviews. A more comprehensive treatment of fairness to both reviewers and papers would require a completely novel approach and is out of the scope of this paper.
Still we must verify that $\GRRR$ and \FairSeq are at least as fair to reviewers as our baselines. For each conference, we compute the distribution of reviewing loads for all algorithms. Our method is relatively consistent with the baselines, and does not introduce a large unfairness in reviewing load. Applying \Cref{alg:pickingreviewerslowerbounds}, we also test $\RRR$ with reviewer load lower bounds of $l_r = 2$ for all $r$ \cite{kobren2019paper}. On all three conferences, the algorithm terminates with complete allocations satisfying reviewer lower and upper bounds, while maintaining EF1 guarantees and competitive USW.

For each algorithm and conference, we compute the number of reviewers receiving each possible reviewing load. These results are displayed in Figures~\ref{fig:rev_loads_midl}, \ref{fig:rev_loads_cvpr}, and \ref{fig:rev_loads_cvpr2018}. These figures also include results from FairIR, another algorithm introduced in \citeA{kobren2019paper} that has not been applied in a real conference setting to date. Reviewers can receive a load in the interval $[0, u_r]$, where $u_r$ is the upper bound for that reviewer. In general, our approaches are about as fair as the other algorithms in terms of reviewer load, though on CVPR there are about 100 reviewers receiving one or two extra papers compared to \FairFlow and \TPMS. 

\begin{table*}
        \caption{Statistics for reviewer-centric scores on all three conferences. Affinities are first transformed by subtracting the maximum affinity (per conference), so that the largest affinity is 0 and the smallest is the negative of the original maximum. USW, NSW, minimum score, and percentile means and standard deviations are computed analogously to the paper-centric statistics. The EF1 requirement is now that there must be some paper reviewer $r$ can drop so that reviewer $r$ has more value than reviewer $r'$. Larger values are better for all statistics besides EF1 violations.}
        \begin{tabu} to \textwidth{X[.9,l]XX[r]X[r]X[r]X[r]X[2,r]X[2,r]}
            \toprule
             & Alg. & USW & NSW & Min Score & EF1 Viol. & Lowest $10$\% & Lowest $25$\% \\
            \midrule
                & \FairFlow & -$.86$ & -$.39$ & -$2.57$ & $6611$ & -$2.34 \pm .16$ & -$2.04 \pm .27$  \\
            & \TPMS & -$.86$ & -$.39$ & -$2.57$ & $6560$ & -$2.34 \pm .15$ & -$2.04 \pm .27$ \\
                MIDL\hspace{1.5em} & \PRFA & -$.88$ & -$.40$ & -$2.60$ & $6632$ & -$2.36 \pm .15$ & -$2.10 \pm .24$ \\
                & $\GRRR$ & -$.88$ & -$.33$ & -$2.86$ & $6667$ & -$2.40 \pm .16$ & -$2.12 \pm .27$ \\
                & $\FairSeqShort$ & -$.88$ & -$.39$ & -$2.77$ & $6659$ & -$2.38 \pm .18$ & -$2.07 \pm .29$  \\

            \midrule
                
                & \FairFlow & -$1.92$ & -$1.15$ & -$6.00$ & $1.645$e$4$ & -$5.50 \pm .50$ & -$4.85 \pm .71$ \\
                 & \TPMS & -$1.77$ & -$.99$ & -$6.00$ & $2.778$e$4$ & -$5.92 \pm .27$ & -$4.91 \pm 1.00$ \\
                CVPR & \PRFA\hspace{2em} & -$1.99$ & -$1.24$ & -$6.00$ & $7.921$e$4$ & -$6.00 \pm 0.00$ & -$5.38 \pm 1.12$ \\
                & $\GRRR$ & -$2.26$ & -$1.50$ & -$6.00$ & $7.865$e$4$ & -$6.00 \pm 0.00$ & -$5.73 \pm .54$ \\
                & $\FairSeqShort$ & -2.05 & -1.33 & -6.00 & 7.468e4 & -$6.00 \pm 0.00$ & -$5.53 \pm .83$ \\
                
            \midrule
    
        \multirow{5}{.5cm}{CVPR '18}  & \FairFlow & -$21.00$ & -$12.26$ & -$95.50$ & $1.119$e$6$ & -$65.50 \pm 11.53$ & -$48.84 \pm 16.07$ \\
          & \TPMS & -$19.72$ & -$11.40$ & -$94.00$ & $1.094$e$6$ & -$63.90 \pm 11.82$ & -$46.93 \pm 16.37$  \\
                &\PRFA\hspace{2em} & -$21.08$ & -$13.10$ & -$93.87$ & $1.134$e$6$ & -$64.01 \pm 11.73$ & -$47.44 \pm 16.08$ \\
                & $\GRRR$ & -$22.21$ & -$12.23$ & -$94.45$ & $1.219$e$6$ & -$69.77 \pm 10.26$ & -$53.28 \pm 15.86$ \\
                & $\FairSeqShort$ & -$21.28$ & -$11.29$ & -$95.22$ & $1.216$e$6$ & -$70.19 \pm 10.48$ & -$52.89 \pm 16.58$ \\

        \end{tabu}
        \label{tab:full_reviewer_stats}
    \end{table*}
We also compared multiple statistics of reviewer welfare and fairness using the affinity scores. Because reviewers consider papers to be chores, we first converted the affinity scores by subtracting the maximum affinity score (per conference) from $v_p(r)$ for all papers $p$ and reviewers $r$. We can then specify $v_r(p)$ using these new values (and bundles are valued additively, as before). The maximum score for any $v_r(p)$ is $0$ and the minimum score is the negative of the original maximum $v_p(r)$. In addition, higher scores are better, with the best score being $0$ either because no papers were assigned or because all assigned papers had maximum affinity. Note that the EF1 criterion is different for chores as well. An allocation $A$ is envy-free up to one paper (EF1) for reviewers if for all pairs of reviewers $r$ and $r'$, $\exists p \in A_r$ such that $v_r(A_r \setminus \{p\}) \geq v_r(A_{r'})$. All statistics prepared this way are presented in Table~\ref{tab:full_reviewer_stats}. Interestingly, it appears that \TPMS has the best efficiency and fairness properties for reviewers, despite the fact that \TPMS especially shows poor fairness properties for papers. No algorithm performs particularly well in terms of reviewer fairness, since all algorithms have fairly large numbers of EF1 violations. These results highlight the tradeoff between fairness for papers and fairness for reviewers. Nothing about these results indicates that obtaining fair outcomes for both papers and reviewers is impossible, but they do highlight the need for an algorithm explicitly designed to be fair to both sides. 


\end{document}

%% file: JAIR/preamble.tex
\usepackage{url}            
\usepackage{booktabs}       
\usepackage{amsmath}
\usepackage{amssymb}
\usepackage{amsthm}
\usepackage{algorithm}
\usepackage[noend]{algorithmic}
\usepackage{graphicx}
\usepackage{thm-restate}
\usepackage{balance} 
\usepackage{tabularx}
\usepackage{tabu}
\usepackage{xspace}
\usepackage{cleveref}
\usepackage{multirow}

\usepackage{subcaption}

\DeclareMathOperator*{\argmax}{arg\,max}
\DeclareMathOperator*{\argmin}{arg\,min}

\DeclareMathOperator*{\USW}{\text{USW}}
\newcommand{\RRR}{\texttt{RRR}}
\newcommand{\GRRR}{\texttt{GRRR}}
\newcommand{\WRP}{\texttt{WRP}}
\newcommand{\FairSeqShort}{\texttt{FairSeq}\xspace}
\newcommand{\FairFlow}{\texttt{FairFlow}\xspace}
\newcommand{\FairSeq}{\texttt{FairSequence}\xspace}
\newcommand{\FairSeqUnchecked}{\texttt{FairSequenceUnchecked}\xspace}
\newcommand{\TPMS}{\texttt{TPMS}\xspace}
\newcommand{\PRFA}{\texttt{PR4A}\xspace}
\newcommand{\alg}{\textrm{alg}}
\newcommand{\possibleset}{{\sc PossibleSet}\xspace}
\newcommand{\possibleutilitarianwelfare}{{\sc PossibleUtilitarianWelfare}\xspace}

\newtheorem{theorem}{Theorem}[section]

\newtheorem{lemma}{Lemma}[section]
\newtheorem{proposition}{Proposition}[section]
\theoremstyle{definition}

\newcommand{\R}{\mathbb{R}}

\renewcommand{\cal}[1]{\mathcal{#1}}

\usepackage{eqparbox}

\renewcommand{\cite}[1]{\shortcite{#1}}
\usepackage{mathtools}
\DeclarePairedDelimiter\abs{\lvert}{\rvert}%

\newcommand{\PapCrit}{f}
\newcommand{\RevCrit}{g}

%% file: main.bbl
\begin{thebibliography}{}

\bibitem[\protect\BCAY{Ahmed, Dickerson,\ \BBA\ Fuge}{Ahmed
  et~al.}{2017}]{ahmed2017diverse}
Ahmed, F., Dickerson, J.~P., \BBA\ Fuge, M. \BBOP2017\BBCP.
\newblock \BBOQ Diverse weighted bipartite b-matching\BBCQ\
\newblock In {\Bem Proceedings of the 26th International Joint Conference on
  Artificial Intelligence (IJCAI)}, \BPGS\ 35--41.

\bibitem[\protect\BCAY{Aziz, Huang, Mattei,\ \BBA\ Segal{-}Halevi}{Aziz
  et~al.}{2019}]{aziz2019constrained}
Aziz, H., Huang, X., Mattei, N., \BBA\ Segal{-}Halevi, E. \BBOP2019\BBCP.
\newblock \BBOQ The constrained round robin algorithm for fair and efficient
  allocation\BBCQ\
\newblock {\Bem CoRR}, {\Bem abs/1908.00161}.

\bibitem[\protect\BCAY{Aziz, Kalinowski, Walsh,\ \BBA\ Xia}{Aziz
  et~al.}{2016}]{aziz2016welfare}
Aziz, H., Kalinowski, T., Walsh, T., \BBA\ Xia, L. \BBOP2016\BBCP.
\newblock \BBOQ Welfare of sequential allocation mechanisms for indivisible
  goods\BBCQ\
\newblock In {\Bem Proceedings of the 22nd European Conference on Artificial
  Intelligence (ECAI)}, \BPGS\ 787--794.

\bibitem[\protect\BCAY{Aziz, Micha,\ \BBA\ Shah}{Aziz
  et~al.}{2023}]{aziz2023group}
Aziz, H., Micha, E., \BBA\ Shah, N. \BBOP2023\BBCP.
\newblock \BBOQ Group fairness in peer review\BBCQ\
\newblock In {\Bem Proceedings of the 22nd International Conference on
  Autonomous Agents and Multi-Agent Systems (AAMAS)}, \BPGS\ 2889--2891.

\bibitem[\protect\BCAY{Aziz, Walsh,\ \BBA\ Xia}{Aziz
  et~al.}{2015}]{aziz2015possible}
Aziz, H., Walsh, T., \BBA\ Xia, L. \BBOP2015\BBCP.
\newblock \BBOQ Possible and necessary allocations via sequential
  mechanisms\BBCQ\
\newblock In {\Bem Proceedings of the 24th International Joint Conference on
  Artificial Intelligence (IJCAI)}, \BPGS\ 468--474.

\bibitem[\protect\BCAY{Babaioff, Ezra,\ \BBA\ Feige}{Babaioff
  et~al.}{2021a}]{babaioff2020fair}
Babaioff, M., Ezra, T., \BBA\ Feige, U. \BBOP2021a\BBCP.
\newblock \BBOQ Fair and truthful mechanisms for dichotomous valuations\BBCQ\
\newblock In {\Bem Proceedings of the 35th AAAI Conference on Artificial
  Intelligence (AAAI)}, \BPGS\ 5119--5126.

\bibitem[\protect\BCAY{Babaioff, Ezra,\ \BBA\ Feige}{Babaioff
  et~al.}{2021b}]{babaioff2021fair}
Babaioff, M., Ezra, T., \BBA\ Feige, U. \BBOP2021b\BBCP.
\newblock \BBOQ Fair-share allocations for agents with arbitrary
  entitlements\BBCQ\
\newblock In {\Bem Proceedings of the 22nd ACM Conference on Economics and
  Computation (EC)}, \BPG\ 127.

\bibitem[\protect\BCAY{Barman, Ghalme, Jain, Kulkarni,\ \BBA\ Narang}{Barman
  et~al.}{2019}]{barman2019fair}
Barman, S., Ghalme, G., Jain, S., Kulkarni, P., \BBA\ Narang, S.
  \BBOP2019\BBCP.
\newblock \BBOQ Fair division of indivisible goods among strategic agents\BBCQ\
\newblock In {\Bem Proceedings of the 18th International Conference on
  Autonomous Agents and Multi-Agent Systems (AAMAS)}, \BPGS\ 1811--1813.

\bibitem[\protect\BCAY{Barman\ \BBA\ Verma}{Barman\ \BBA\
  Verma}{2021}]{barman2021existence}
Barman, S.\BBACOMMA\  \BBA\ Verma, P. \BBOP2021\BBCP.
\newblock \BBOQ Existence and computation of maximin fair allocations under
  matroid-rank valuations\BBCQ\
\newblock In {\Bem Proceedings of the 20th International Conference on
  Autonomous Agents and Multi-Agent Systems (AAMAS)}, \BPGS\ 169--177.

\bibitem[\protect\BCAY{Benabbou, Chakraborty, Igarashi,\ \BBA\ Zick}{Benabbou
  et~al.}{2020}]{benabbou2020finding}
Benabbou, N., Chakraborty, M., Igarashi, A., \BBA\ Zick, Y. \BBOP2020\BBCP.
\newblock \BBOQ Finding fair and efficient allocations when valuations don't
  add up\BBCQ\
\newblock In {\Bem Proceedings of the 13th International Symposium on
  Algorithmic Game Theory (SAGT)}, \BPGS\ 32--46.

\bibitem[\protect\BCAY{Biswas\ \BBA\ Barman}{Biswas\ \BBA\
  Barman}{2018}]{biswas2018fair}
Biswas, A.\BBACOMMA\  \BBA\ Barman, S. \BBOP2018\BBCP.
\newblock \BBOQ Fair division under cardinality constraints.\BBCQ\
\newblock In {\Bem Proceedings of the 27th International Joint Conference on
  Artificial Intelligence (IJCAI)}, \BPGS\ 91--97.

\bibitem[\protect\BCAY{Bouveret, Chevaleyre,\ \BBA\ Maudet}{Bouveret
  et~al.}{2016}]{brandt2016handbookfairdiv}
Bouveret, S., Chevaleyre, Y., \BBA\ Maudet, N. \BBOP2016\BBCP.
\newblock \BBOQ Fair allocation of indivisible goods\BBCQ\
\newblock In Brandt, F., Conitzer, V., Endriss, U., Lang, J., \BBA\ Procaccia,
  A.~D.\BEDS, {\Bem Handbook of computational social choice}, \BPGS\ 284--309.
  Cambridge University Press.

\bibitem[\protect\BCAY{Bouveret\ \BBA\ Lang}{Bouveret\ \BBA\
  Lang}{2011}]{bouveret2011general}
Bouveret, S.\BBACOMMA\  \BBA\ Lang, J. \BBOP2011\BBCP.
\newblock \BBOQ A general elicitation-free protocol for allocating indivisible
  goods\BBCQ\
\newblock In {\Bem Proceedings of the 22nd International Joint Conference on
  Artificial Intelligence (IJCAI)}, \BPGS\ 73--78.

\bibitem[\protect\BCAY{Buchbinder, Feldman, Naor,\ \BBA\ Schwartz}{Buchbinder
  et~al.}{2012}]{buchbinder2012tight}
Buchbinder, N., Feldman, M., Naor, J., \BBA\ Schwartz, R. \BBOP2012\BBCP.
\newblock \BBOQ A tight linear time (1/2)-approximation for unconstrained
  submodular maximization\BBCQ\
\newblock In {\Bem Proceedings of the 53rd Symposium on Foundations of Computer
  Science (FOCS)}, \BPGS\ 649--658.

\bibitem[\protect\BCAY{Budish}{Budish}{2011}]{budish2011ef1}
Budish, E. \BBOP2011\BBCP.
\newblock \BBOQ The combinatorial assignment problem: Approximate competitive
  equilibrium from equal incomes\BBCQ\
\newblock {\Bem Journal of Political Economy}, {\Bem 119}, 1061--1103.

\bibitem[\protect\BCAY{Caragiannis, Kurokawa, Moulin, Procaccia, Shah,\ \BBA\
  Wang}{Caragiannis et~al.}{2019}]{caragiannis2019unreasonable}
Caragiannis, I., Kurokawa, D., Moulin, H., Procaccia, A.~D., Shah, N., \BBA\
  Wang, J. \BBOP2019\BBCP.
\newblock \BBOQ The unreasonable fairness of maximum {Nash} welfare\BBCQ\
\newblock {\Bem ACM Transactions on Economics and Computation (TEAC)}, {\Bem
  7\/}(3), 1--32.

\bibitem[\protect\BCAY{Caragiannis\ \BBA\ Rathi}{Caragiannis\ \BBA\
  Rathi}{2023}]{caragiannis2023optimizing}
Caragiannis, I.\BBACOMMA\  \BBA\ Rathi, N. \BBOP2023\BBCP.
\newblock \BBOQ Optimizing over serial dictatorships\BBCQ\
\newblock In {\Bem Proceedings of the 16th International Symposium on
  Algorithmic Game Theory (SAGT)}, \BPGS\ 329--346.

\bibitem[\protect\BCAY{Chakraborty, Igarashi, Suksompong,\ \BBA\
  Zick}{Chakraborty et~al.}{2021a}]{wef1}
Chakraborty, M., Igarashi, A., Suksompong, W., \BBA\ Zick, Y. \BBOP2021a\BBCP.
\newblock \BBOQ Weighted envy-freeness in indivisible item allocation\BBCQ\
\newblock {\Bem ACM Trans. Econ. Comput.}, {\Bem 9\/}(3), 1--39.

\bibitem[\protect\BCAY{Chakraborty, Schmidt-Kraepelin,\ \BBA\
  Suksompong}{Chakraborty et~al.}{2021b}]{CHAKRABORTY2021103578}
Chakraborty, M., Schmidt-Kraepelin, U., \BBA\ Suksompong, W. \BBOP2021b\BBCP.
\newblock \BBOQ Picking sequences and monotonicity in weighted fair
  division\BBCQ\
\newblock {\Bem Artificial Intelligence}, {\Bem 301}.

\bibitem[\protect\BCAY{Chakraborty, Segal-Halevi,\ \BBA\
  Suksompong}{Chakraborty et~al.}{2022}]{chakraborty2022weighted}
Chakraborty, M., Segal-Halevi, E., \BBA\ Suksompong, W. \BBOP2022\BBCP.
\newblock \BBOQ Weighted fairness notions for indivisible items revisited\BBCQ\
\newblock In {\Bem Proceedings of the 36th AAAI Conference on Artificial
  Intelligence (AAAI)}, \BPGS\ 4949--4956.

\bibitem[\protect\BCAY{Charlin\ \BBA\ Zemel}{Charlin\ \BBA\
  Zemel}{2013}]{charlin2013toronto}
Charlin, L.\BBACOMMA\  \BBA\ Zemel, R. \BBOP2013\BBCP.
\newblock \BBOQ The {T}oronto paper matching system: An automated
  paper-reviewer assignment system\BBCQ\
\newblock In {\Bem Proceedings of the 2013 ICML Workshop on Peer Reviewing and
  Publishing Models}.

\bibitem[\protect\BCAY{Charlin, Zemel,\ \BBA\ Boutilier}{Charlin
  et~al.}{2011}]{charlin2011framework}
Charlin, L., Zemel, R., \BBA\ Boutilier, C. \BBOP2011\BBCP.
\newblock \BBOQ A framework for optimizing paper matching\BBCQ\
\newblock In {\Bem Proceedings of the 27th Annual Conference on Uncertainty in
  Artificial Intelligence (UAI)}, \BPGS\ 86--95.

\bibitem[\protect\BCAY{Conry, Koren,\ \BBA\ Ramakrishnan}{Conry
  et~al.}{2009}]{conry2009recommender}
Conry, D., Koren, Y., \BBA\ Ramakrishnan, N. \BBOP2009\BBCP.
\newblock \BBOQ Recommender systems for the conference paper assignment
  problem\BBCQ\
\newblock In {\Bem Proceedings of the 3rd ACM Conference on Recommendation
  Systems (RecSys)}, \BPGS\ 357--360.

\bibitem[\protect\BCAY{Cousins, Payan,\ \BBA\ Zick}{Cousins
  et~al.}{2023a}]{cousins2023into}
Cousins, C., Payan, J., \BBA\ Zick, Y. \BBOP2023a\BBCP.
\newblock \BBOQ Into the unknown: Assigning reviewers to papers with uncertain
  affinities\BBCQ\
\newblock In {\Bem Proceedings of the 16th International Symposium on
  Algorithmic Game Theory (SAGT)}, \BPGS\ 179--197.

\bibitem[\protect\BCAY{Cousins, Viswanathan,\ \BBA\ Zick}{Cousins
  et~al.}{2023b}]{cousins2023dividing}
Cousins, C., Viswanathan, V., \BBA\ Zick, Y. \BBOP2023b\BBCP.
\newblock \BBOQ Dividing good and great items among agents with bivalued
  submodular valuations\BBCQ\
\newblock In {\Bem Proceedings of the 19th International Conference on Web and
  Internet Economics (WINE)}, \BPGS\ 225--241.

\bibitem[\protect\BCAY{Cousins, Viswanathan,\ \BBA\ Zick}{Cousins
  et~al.}{2023c}]{cousins2023good}
Cousins, C., Viswanathan, V., \BBA\ Zick, Y. \BBOP2023c\BBCP.
\newblock \BBOQ The good, the bad and the submodular: Fairly allocating mixed
  manna under order-neutral submodular preferences\BBCQ\
\newblock In {\Bem Proceedings of the 19th International Conference on Web and
  Internet Economics (WINE)}, \BPGS\ 207--224.

\bibitem[\protect\BCAY{Das\ \BBA\ Kempe}{Das\ \BBA\
  Kempe}{2011}]{das2011submodular}
Das, A.\BBACOMMA\  \BBA\ Kempe, D. \BBOP2011\BBCP.
\newblock \BBOQ Submodular meets spectral: greedy algorithms for subset
  selection, sparse approximation and dictionary selection\BBCQ\
\newblock In {\Bem Proceedings of the 28th International Conference on Machine
  Learning (ICML)}, \BPGS\ 1057--1064.

\bibitem[\protect\BCAY{Dror, Feldman,\ \BBA\ Segal-Halevi}{Dror
  et~al.}{2021}]{dror2021fair}
Dror, A., Feldman, M., \BBA\ Segal-Halevi, E. \BBOP2021\BBCP.
\newblock \BBOQ On fair division under heterogeneous matroid constraints\BBCQ\
\newblock In {\Bem Proceedings of the 35th AAAI Conference on Artificial
  Intelligence (AAAI)}, \BPGS\ 5312--5320.

\bibitem[\protect\BCAY{Fisher, Nemhauser,\ \BBA\ Wolsey}{Fisher
  et~al.}{1978}]{fisher1978analysis}
Fisher, M.~L., Nemhauser, G.~L., \BBA\ Wolsey, L.~A. \BBOP1978\BBCP.
\newblock \BBOQ An analysis of approximations for maximizing submodular set
  functions---{II}\BBCQ\
\newblock In Balinski, M.~L.\BBACOMMA\  \BBA\ Hoffman, A.~J.\BEDS, {\Bem
  Polyhedral Combinatorics}, \BPGS\ 73--87. Springer Berlin Heidelberg.

\bibitem[\protect\BCAY{Garg, Kavitha, Kumar, Mehlhorn,\ \BBA\ Mestre}{Garg
  et~al.}{2010}]{garg2010assigning}
Garg, N., Kavitha, T., Kumar, A., Mehlhorn, K., \BBA\ Mestre, J.
  \BBOP2010\BBCP.
\newblock \BBOQ Assigning papers to referees\BBCQ\
\newblock {\Bem Algorithmica}, {\Bem 58\/}(1), 119--136.

\bibitem[\protect\BCAY{Gini}{Gini}{1936}]{gini1936measure}
Gini, C. \BBOP1936\BBCP.
\newblock \BBOQ On the measure of concentration with special reference to
  income and statistics\BBCQ\
\newblock {\Bem Colorado College Publication, General Series}, {\Bem 208\/}(1),
  73--79.

\bibitem[\protect\BCAY{G{\"o}lz\ \BBA\ Procaccia}{G{\"o}lz\ \BBA\
  Procaccia}{2019}]{golz2019migration}
G{\"o}lz, P.\BBACOMMA\  \BBA\ Procaccia, A.~D. \BBOP2019\BBCP.
\newblock \BBOQ Migration as submodular optimization\BBCQ\
\newblock In {\Bem Proceedings of the 33rd AAAI Conference on Artificial
  Intelligence (AAAI)}, \BPGS\ 549--556.

\bibitem[\protect\BCAY{Hartvigsen, Wei,\ \BBA\ Czuchlewski}{Hartvigsen
  et~al.}{1999}]{hartvigsen1999conference}
Hartvigsen, D., Wei, J.~C., \BBA\ Czuchlewski, R. \BBOP1999\BBCP.
\newblock \BBOQ The conference paper-reviewer assignment problem\BBCQ\
\newblock {\Bem Decision Sciences}, {\Bem 30\/}(3), 865--876.

\bibitem[\protect\BCAY{Jecmen, Zhang, Liu, Shah, Conitzer,\ \BBA\ Fang}{Jecmen
  et~al.}{2020}]{jecmen2020mitigating}
Jecmen, S., Zhang, H., Liu, R., Shah, N., Conitzer, V., \BBA\ Fang, F.
  \BBOP2020\BBCP.
\newblock \BBOQ Mitigating manipulation in peer review via randomized reviewer
  assignments\BBCQ\
\newblock In {\Bem Proceedings of the 33rd Annual Conference on Neural
  Information Processing Systems (NeurIPS)}, \BPGS\ 12533--12545.

\bibitem[\protect\BCAY{Kalinowski, Narodytska,\ \BBA\ Walsh}{Kalinowski
  et~al.}{2013}]{kalinowski2013social}
Kalinowski, T., Narodytska, N., \BBA\ Walsh, T. \BBOP2013\BBCP.
\newblock \BBOQ A social welfare optimal sequential allocation procedure\BBCQ\
\newblock In {\Bem Proceedings of the 23rd International Joint Conference on
  Artificial Intelligence (IJCAI)}, \BPGS\ 227--233.

\bibitem[\protect\BCAY{Kobren, Saha,\ \BBA\ McCallum}{Kobren
  et~al.}{2019}]{kobren2019paper}
Kobren, A., Saha, B., \BBA\ McCallum, A. \BBOP2019\BBCP.
\newblock \BBOQ Paper matching with local fairness constraints\BBCQ\
\newblock In {\Bem Proceedings of the 25th International Conference on
  Knowledge Discovery and Data Mining (KDD)}, \BPGS\ 1247--1257.

\bibitem[\protect\BCAY{Kou, U, Mamoulis,\ \BBA\ Gong}{Kou
  et~al.}{2015}]{kou2015weighted}
Kou, N.~M., U, L.~H., Mamoulis, N., \BBA\ Gong, Z. \BBOP2015\BBCP.
\newblock \BBOQ Weighted coverage based reviewer assignment\BBCQ\
\newblock In {\Bem Proceedings of the 2015 ACM SIGMOD International Conference
  on Management of Data}, \BPGS\ 2031--2046.

\bibitem[\protect\BCAY{Leyton-Brown, Nandwani, Zarkoob, Cameron, Newman, Raghu,
  et~al.}{Leyton-Brown et~al.}{2022}]{leyton2022matching}
Leyton-Brown, K., Nandwani, Y., Zarkoob, H., Cameron, C., Newman, N., Raghu,
  D., et~al. \BBOP2022\BBCP.
\newblock \BBOQ Matching papers and reviewers at large conferences\BBCQ\
\newblock {\Bem CoRR}, {\Bem abs/2202.12273}.

\bibitem[\protect\BCAY{Lian, Mattei, Noble,\ \BBA\ Walsh}{Lian
  et~al.}{2018}]{lian2018conference}
Lian, J.~W., Mattei, N., Noble, R., \BBA\ Walsh, T. \BBOP2018\BBCP.
\newblock \BBOQ The conference paper assignment problem: Using order weighted
  averages to assign indivisible goods\BBCQ\
\newblock In {\Bem Proceedings of the 32nd AAAI Conference on Artificial
  Intelligence (AAAI)}, \BPGS\ 1138--1145.

\bibitem[\protect\BCAY{Lipton, Markakis, Mossel,\ \BBA\ Saberi}{Lipton
  et~al.}{2004}]{lipton2004approximately}
Lipton, R.~J., Markakis, E., Mossel, E., \BBA\ Saberi, A. \BBOP2004\BBCP.
\newblock \BBOQ On approximately fair allocations of indivisible goods\BBCQ\
\newblock In {\Bem Proceedings of the 5th ACM Conference on Electronic Commerce
  (EC)}, \BPGS\ 125--131.

\bibitem[\protect\BCAY{Long, Wong, Peng,\ \BBA\ Ye}{Long
  et~al.}{2013}]{long2013good}
Long, C., Wong, R. C.-W., Peng, Y., \BBA\ Ye, L. \BBOP2013\BBCP.
\newblock \BBOQ On good and fair paper-reviewer assignment\BBCQ\
\newblock In {\Bem Proceedings of the 13th IEEE International Conference on
  Data Mining (ICDM)}, \BPGS\ 1145--1150.

\bibitem[\protect\BCAY{Montanari, Schmidt-Kraepelin, Suksompong,\ \BBA\
  Teh}{Montanari et~al.}{2024}]{montanari2024weighted}
Montanari, L., Schmidt-Kraepelin, U., Suksompong, W., \BBA\ Teh, N.
  \BBOP2024\BBCP.
\newblock \BBOQ Weighted envy-freeness for submodular valuations\BBCQ\
\newblock In {\Bem Proceedings of the 38th AAAI Conference on Artificial
  Intelligence (AAAI)}.

\bibitem[\protect\BCAY{Mulligan, Hall,\ \BBA\ Raphael}{Mulligan
  et~al.}{2013}]{mulligan2013peer}
Mulligan, A., Hall, L., \BBA\ Raphael, E. \BBOP2013\BBCP.
\newblock \BBOQ Peer review in a changing world: An international study
  measuring the attitudes of researchers\BBCQ\
\newblock {\Bem Journal of the American Society for Information Science and
  Technology}, {\Bem 64\/}(1), 132--161.

\bibitem[\protect\BCAY{Mysore, Jasim, McCallum,\ \BBA\ Zamani}{Mysore
  et~al.}{2023}]{mysore2023editable}
Mysore, S., Jasim, M., McCallum, A., \BBA\ Zamani, H. \BBOP2023\BBCP.
\newblock \BBOQ Editable user profiles for controllable text
  recommendation\BBCQ\
\newblock In {\Bem Proceedings of the 46th International ACM SIGIR Conference
  on Research and Development in Information Retrieval (SIGIR)}, \BPGS\
  993--1003.

\bibitem[\protect\BCAY{O'Dell, Wattenhofer,\ \BBA\ Wattenhofer}{O'Dell
  et~al.}{2005}]{odell2005paper}
O'Dell, R., Wattenhofer, M., \BBA\ Wattenhofer, R. \BBOP2005\BBCP.
\newblock \BBOQ The paper assignment problem\BBCQ\
\newblock \BTR\ 491, Department of Computer Science, ETH Zurich.

\bibitem[\protect\BCAY{Oxley}{Oxley}{2011}]{oxley2011matroid}
Oxley, J. \BBOP2011\BBCP.
\newblock {\Bem Matroid Theory\/} (2nd \BEd).
\newblock Oxford Univerity Press.

\bibitem[\protect\BCAY{Payan\ \BBA\ Zick}{Payan\ \BBA\
  Zick}{2022}]{payan2022will}
Payan, J.\BBACOMMA\  \BBA\ Zick, Y. \BBOP2022\BBCP.
\newblock \BBOQ I will have order! {O}ptimizing orders for fair reviewer
  assignment\BBCQ\
\newblock In {\Bem Proceedings of the 31st International Joint Conference on
  Artificial Intelligence (IJCAI)}, \BPGS\ 440--446.

\bibitem[\protect\BCAY{Rozenzweig, Meir, Mattei,\ \BBA\ Amir}{Rozenzweig
  et~al.}{2023}]{rozenzweig2023mitigating}
Rozenzweig, I., Meir, R., Mattei, N., \BBA\ Amir, O. \BBOP2023\BBCP.
\newblock \BBOQ Mitigating skewed bidding for conference paper assignment\BBCQ\
\newblock In {\Bem Proceedings of the 22nd International Conference on
  Autonomous Agents and Multi-Agent Systems (AAMAS)}, \BPGS\ 573--581.

\bibitem[\protect\BCAY{Shah}{Shah}{2019}]{shah2019principled}
Shah, N.~B. \BBOP2019\BBCP.
\newblock \BBOQ Principled methods to improve peer review\BBCQ\
\newblock
  \url{https://www.cs.cmu.edu/~nihars/publications/survey_peerreview_niharshah.pdf}.
\newblock Retrieved on February 14, 2024.

\bibitem[\protect\BCAY{Shah}{Shah}{2022}]{shah2022challenges}
Shah, N.~B. \BBOP2022\BBCP.
\newblock \BBOQ Challenges, experiments, and computational solutions in peer
  review\BBCQ\
\newblock {\Bem Communications of the ACM}, {\Bem 65\/}(6), 76--87.

\bibitem[\protect\BCAY{Stelmakh}{Stelmakh}{2021}]{stelmakh2021towards}
Stelmakh, I. \BBOP2021\BBCP.
\newblock \BBOQ Towards fair, equitable, and efficient peer review\BBCQ\
\newblock In {\Bem Proceedings of the 35th AAAI Conference on Artificial
  Intelligence (AAAI)}, \BPGS\ 15736--15737.

\bibitem[\protect\BCAY{Stelmakh, Shah,\ \BBA\ Singh}{Stelmakh
  et~al.}{2019}]{stelmakh2019peerreview4all}
Stelmakh, I., Shah, N.~B., \BBA\ Singh, A. \BBOP2019\BBCP.
\newblock \BBOQ {PeerReview4All}: Fair and accurate reviewer assignment in peer
  review\BBCQ\
\newblock In {\Bem Proceedings of the 30th International Conference on
  Algorithmic Learning Theory (ALT)}, \BPGS\ 828--856.

\bibitem[\protect\BCAY{Suksompong\ \BBA\ Teh}{Suksompong\ \BBA\
  Teh}{2022}]{suksompong2022maximum}
Suksompong, W.\BBACOMMA\  \BBA\ Teh, N. \BBOP2022\BBCP.
\newblock \BBOQ On maximum weighted nash welfare for binary valuations\BBCQ\
\newblock {\Bem Mathematical Social Sciences}, {\Bem 117}, 101--108.

\bibitem[\protect\BCAY{Suksompong\ \BBA\ Teh}{Suksompong\ \BBA\
  Teh}{2023}]{SUKSOMPONG202348}
Suksompong, W.\BBACOMMA\  \BBA\ Teh, N. \BBOP2023\BBCP.
\newblock \BBOQ Weighted fair division with matroid-rank valuations:
  Monotonicity and strategyproofness\BBCQ\
\newblock {\Bem Mathematical Social Sciences}, {\Bem 126}, 48--59.

\bibitem[\protect\BCAY{Tan, Dai, Ren, Walsh,\ \BBA\ Aleksandrov}{Tan
  et~al.}{2021}]{tan2021minimal}
Tan, M., Dai, Z., Ren, Y., Walsh, T., \BBA\ Aleksandrov, M. \BBOP2021\BBCP.
\newblock \BBOQ Minimal-envy conference paper assignment: {F}ormulation and a
  fast iterative algorithm\BBCQ\
\newblock In {\Bem Proceedings of the 5th Asian Conference on Artificial
  Intelligence Technology (ACAIT)}, \BPGS\ 667--674.

\bibitem[\protect\BCAY{Viswanathan\ \BBA\ Zick}{Viswanathan\ \BBA\
  Zick}{2023a}]{viswanathan2023general}
Viswanathan, V.\BBACOMMA\  \BBA\ Zick, Y. \BBOP2023a\BBCP.
\newblock \BBOQ A general framework for fair allocation under matroid rank
  valuations\BBCQ\
\newblock In {\Bem Proceedings of the 24th ACM Conference on Economics and
  Computation (EC)}, \BPGS\ 1129--1152.

\bibitem[\protect\BCAY{Viswanathan\ \BBA\ Zick}{Viswanathan\ \BBA\
  Zick}{2023b}]{viswanathan2023yankee}
Viswanathan, V.\BBACOMMA\  \BBA\ Zick, Y. \BBOP2023b\BBCP.
\newblock \BBOQ Yankee swap: A fast and simple fair allocation mechanism for
  matroid rank valuations\BBCQ\
\newblock In {\Bem Proceedings of the 22nd International Conference on
  Autonomous Agents and Multi-Agent Systems (AAMAS)}, \BPGS\ 179--187.

\end{thebibliography}
